# On the Suitability of Blockchain Platforms for IoT Applications: Architectures, Security, Privacy, and Performance⋆,⋆⋆

Sotirios **Brotsis**[a], Konstantinos **Limniotis**[a,b], Gueltoum **Bendiab**[c], Nicholas **Kolokotronis**[a] and Stavros **Shiaeles**[c]

[a]*University of Peloponnese, Department of Informatics and Telecommunications, 22131 Tripolis, Greece*
[b]*Hellenic Data Protection Authority, 11523 Athens, Greece*
[c]*University of Portsmouth, Cyber Security Research Group, PO1 2UP, Portsmouth, UK*

**ARTICLE INFO**

*Keywords*:
Blockchain
Consensus protocols
Cyber-attacks
Fault tolerance
Internet of things
Security
Smart contracts
Smart homes
Performance
Privacy

**ABSTRACT**

Blockchain and distributed ledger technologies have received significant interest in various areas beyond the financial sector, with profound applications in the Internet of Things (IoT), providing the means for creating truly trustless and secure solutions for IoT applications. Taking into account the weak security defences that the majority of IoT devices have, it is critical that a blockchain-based solution targeting the IoT is not only capable of addressing the many challenges IoT is facing, but also does not introduce other defects, e.g. in terms of performance, making its adoption hard to achieve. This paper aims at addressing the above needs by providing a comprehensive and coherent review of the available blockchain solutions to determine their ability to meet the requirements and tackle the challenges of the IoT, using the smart home as the reference domain. Key architectural aspects of blockchain solutions, like the platforms' software and network setups, the consensus protocols used, as well as smart contracts, are examined in terms of their ability to withstand various types of common IoT and blockchain attacks, deliver enhanced privacy features, and assure adequate performance levels while processing large amounts of transactions being generated in an IoT environment. The analysis carried out identified that the defences currently provided by blockchain platforms are not sufficient to thwart all the prominent attacks against blockchains, with blockchain 1.0 and 2.0 platforms being susceptible to the majority of them. On the other side, privacy related mechanisms are being supported, to varying degrees, by all platforms investigated; however, each of the them tackles specific only privacy aspects, thus rendering the overall privacy evaluation a challenging task which needs to be considered in an ad-hoc basis. If the underlying consensus protocols' performance and fault tolerance is also considered, then only a small number of platforms meet the requirements of our reference IoT domain.

## 1. Introduction

The impact of blockchain technology has already begun to be widely known and this may just be the tip of the iceberg. Since the evolution of Bitcoin, the virtual currency created by Satoshi Nakamoto [165], the blockchain technology has met growing interest in the last years as a novel technology facilitating the degree of decentralisation required by modern applications and services in an efficient and robust way. In simple words, blockchain is a distributed database of records, or shared ledger of all the transactions or digital events having been executed and exchanged among a number of parties. The key characteristic of this distributed database is the fact that each transaction in the ledger has been verified by the majority of the network's peers (via a consensus protocol) and is immutable, therefore inherently supporting transparency and considerably high security level.

Apart from the financial sector and the development of decentralised currency systems, the IoT constitutes a prominent application domain of blockchain with a large number of frameworks, architectures, and solutions having been proposed so far to improve the operational characteristics and overcome the limitations of the current systems [110, 157]. Application domains range from smart grid, smart cities and smart homes to IoT cloud and IoT edge following the recent advancements in telecommunications to improve latency and other network performance metrics. In particular, the adoption of blockchain in the IoT has already led to powerful decentralised information systems, allowing IoT devices to act autonomously and carry out transactions by executing smart contracts [46]. At the same time, the technological evolution brought by the IoT gives birth to new forms of cyber-attacks that take advantage of the heterogeneity and complexity of the IoT ecosystem. Current centralised security defences, apart from the fact that they often rely on trusted third parties, they are inadequate to deal with sophisticated attacks that exploit the existence of various vulnerabilities in IoT devices and platforms. Therefore, applications of blockchain towards enhancing the security and privacy of the IoT have received great attention due to its capacity in providing a fundamentally different approach as a distributed, transparent and accountable sys-This project has received funding from the European Union's Horizon 2020 research and innovation programme under grant agreement no. 786698. The work reflects only the authors' view and the Agency is not responsible for any use that may be made of the information it contains.

✉ brotsis@uop.gr (S. Brotsis); klimn@uop.gr (K. Limniotis); gueltoum.bendiab@port.ac.uk (G. Bendiab); nkolok@uop.gr (N. Kolokotronis); stavros.shiaeles@port.ac.uk (S. Shiaeles)

ORCID(s): 0000-0002-6053-7980 (S. Brotsis); 0000-0002-7663-7169 (K. Limniotis); 0000-0002-9843-5496 (G. Bendiab); 0000-0003-0660-8431 (N. Kolokotronis); 0000-0003-3866-0672 (S. Shiaeles)

tem [54, 100, 169]. Examples of application areas include identity security [24], intrusion detection systems and mitigation of distributed denial-of-service attacks [15, 99], data security and digital forensic evidence preservation [2, 38], as well as privacy enhancement and management [70, 102]. Specifically for the IoT, the security through transparency approach in public blockchains has in many cases clear advantages compared to the usual security through obscurity model. However, there are major concerns when integrating blockchain with the IoT [147, 159, 170]. The blockchain network needs to: (a) support an increasingly huge number of IoT devices, which could join or leave the network at any time; (b) withstand adversarial behaviour and mitigate the attacks occurring in an IoT environment; (c) be explicitly designed for the IoT ecosystem, taking into account the limited resources (computing power, storage, memory) available in IoT devices; (d) achieve high performance; and (e) satisfy the needs for increased user privacy.

In this article we perform a systematic literature review of a large number of dominant blockchain platforms and their consensus protocols to define their ability to meet the requirements and tackle the challenges imposed by IoT applications. In particular, starting from our preliminary results in [37], which considered the security aspects of a limited number of consensus protocols, we considerably extend our study in terms of both the blockchain platforms and consensus protocols investigated but also on the criteria against which these are evaluated. Our reference IoT domain is that of a smart home or *small-office, home-office* (SOHO), due to the fact that it constitutes a prominent application domain of the IoT platforms and services, thus extracting the requirements that the blockchain solutions should meet. Compared to previous relative surveys of blockchain-enabled IoT applications [5, 110, 164], in which a limited number of blockchain protocols and/or platforms is compared, little or no focus on security and privacy is given, whereas the suitability of blockchain implementations to the IoT is investigated with unclear evaluation criteria, we provide a holistic approach with a large number of factors considered jointly on a common ground and a reference IoT use case (i.e. the smart home ecosystem). More precisely, we utilise seven critical factors to evaluate in a quantitative manner (instead of a qualitative one), following an extensive analysis, the suitability of 32 well-known blockchain platforms and the underlying 16 consensus protocols for use in IoT applications along with their ability to address various IoT challenges; these factors fall into the broad areas of security, privacy, and performance. The high-level contributions of this work are summarised as follows:

- Presentation, analysis, and comparative evaluation of a *large number of dominant* blockchain platforms and consensus protocols under a common framework and working assumptions linked to a concrete IoT domain with realistic key performance indicators;
- Evaluation is *quantitative* covering the broad areas of security, privacy, and performance (they constitute the grand challenges of the IoT) and is supported by an *extensive* analysis of the relative aspects to allow identifying an optimal balance between the above;
- Consideration of security issues at *all layers* of a blockchain architecture instead of the consensus protocols only — a detailed investigation is carried out to identify the availability of defences against state-of-the-art high-impact attacks towards blockchains;
- Identification of the prominent and emerging privacy-enhancing techniques in blockchains and investigation of the extent to which privacy would be safeguarded in the blockchain platforms and consensus protocols of interest if integrated in IoT applications;
- Analysis of the performance and security trade-off in a large number of blockchain consensus protocols with the goal of supporting Byzantine fault tolerance due to the adversarial nature of the IoT ecosystem.

In terms of security, our analysis indicates that none of the blockchain platforms considered in this work has proved to be successful in addressing all the cyber-attacks; to be more precise, most of the platforms provide security defences to mitigate only some of the attacks, with blockchain 1.0 and 2.0 platforms being susceptible to the majority of them (thus making their use in security-critical IoT applications rather questionable). Hedera [19] was an exception, as it provides protection against most of the considered attacks. In terms of privacy, it was seen that all of the blockchain platforms considered provide some privacy preservation mechanisms, at a minimum supporting the features of pseudonymisation and off-chain storage (in case of private permissioned blockchain). Depending on the intended usage (related to cross-industry usage), some platforms such as Ethereum, Hyperledger Fabric, Quorum, have shown to provide an increased support for privacy-enhancing mechanisms compared to the others. However, we show that none privacy-enhancing mechanism should be considered as a panacea per se. Finally, it was shown that if platforms' security and privacy is jointly considered with the underlying consensus protocols' performance and fault tolerance, then only a small number of cases meet the requirements of our reference IoT domain. New and emerging protocols, such as BFT-SMaRt, PoET, VBFT, Exonum, and DAG-based consensus, seem to satisfy these requirements if implemented in a secure, permissioned and high performance blockchain platform.

The rest of the paper is organised as follows. Section 2 provides an overview on blockchain technology and Internet of Things (IoT) challenges. In section 3, we investigate a prominent IoT domain, namely the case of smart homes, and how various proposed solutions utilising blockchain are being used to tackle security, privacy, and other IoT challenges, while the requirements imposed on blockchain solutions by the domain under consideration are discussed in section 4. Section 5 provides an brief introduction to fault tolerance mechanisms that are commonly used in distributed systems, whereas their incorporation into various blockchain platforms' consensus protocols (16 protocol families) is further analysed in detail in section 6, in conjunction with their ability to meet the requirements identified for the IoT. A thor-ough investigation of the security properties of a number of well-known (32 in total) blockchain platforms is being provided in section 7, focusing on their ability to thwart common types of attacks on blockchains (covering areas ranging from network attacks to cryptographic attacks), while the smart contracts' considerations, security and verification are treated in section 8; likewise, an extensive analysis on privacy aspects and mechanisms provided to comply with the *General Data Protection Directive* (GDPR) is given in section 9. Finally, a comparative evaluation of blockchain platforms is carried out in section 10 to determine the degree of their suitability in IoT (smart home) applications; the conclusions of the paper are provided in section 11.

## 2. Background and related work

### 2.1. IoT considerations and challenges

The proliferation of the wireless communication in digital technologies has led to the ongoing technological revolution of the IoT. The cyber-physical devices, such as smart watches, laptops, tablets, smart televisions, patients or infants monitoring systems and medical implants, have become an indispensable part of our today's society, with the IoT technology being implemented into various fields. However, these IoT devices appeared to be an easy victim for the attackers due to their flaws, mainly stemming from the inherent limitations of the IoT devices. More precisely, device manufacturers put emphasis on the devices' cost, size and usability, without taking into account the requirements for power consumption and computational capabilities that are needed to implement robust security mechanisms. Hence, lightweight security practices are being usually adopted, only because a possible exploitation of some of their IoT products might result to the loss of the company's reputation.

Therefore, to counter system flaws, cloud computing has been proposed to enhance the IoT applications through providing significant security and privacy preservation mechanisms [24, 178]. Unfortunately, these implementations are based on centralised systems that are associated with cloud servers and a *trusted-third-party* (TTP) to address the necessary requirements of an IoT ecosystem. Although such solutions might be applicable for the time being, various barriers still exist. With the significantly increasing number of the IoT devices, there is a need for more decentralised networks to cope with the aforementioned issues. Therefore, a synopsis of the most prominent challenges that the IoT implementations face is being provided below:

- *Security*: The design and provisioning of a secure solution is a challenging task. Flawed designs and implementations allow cyber-attackers to easily compromise IoT devices and use them as the means for launching other advanced attacks, such as the distributed denial of service (DDoS) attack against Dyn that was attributed to Mirai malware. Current implementations, in which a private cloud acts as single point of failure, can wreak havoc to the network's security, if it is compromised. Even if a single device is attacked, malicious messages and malware can be sent to other devices, which in turn can result to the exploitation of the device or even to the manipulation of the entire information distributed within the IoT ecosystem.
- *Privacy*: With the ever increasing number of the IoT devices and the huge amounts of the generated data, privacy has become a critical concern in the IoT ecosystem and therefore, the device owner needs to be assured that there will be no leakage when its private information is being processed. Unfortunately, many companies allow access to third parties (such as internet service providers) without providing adequate or precise information to their customers of what has been disclosed and for which purposes [133]. Possible leakage or misuse of personal data in an IoT environment can result to significant privacy violations. Therefore, secure mechanisms for ensuring lawful, fair and transparent personal data processing, without disclosing data to unauthorised entities or using the data for other purposes than the original one, are necessary.
- *Encryption*: Before the transmission of the IoT data to the storage mechanism, it is required to be encrypted. The heterogeneity and the limited computation power that the IoT devices posses make the encryption process a significant challenge, since most of the secure cryptographic algorithms (such as the ECDSA) need a significant amount of computational power and energy. Some devices may employ security network protocols, such as IPsec, SSL/TLS and DTL. Due to the existing inherent limitations, such protocols can only be implemented in devices like the gateways that do not suffer, at this level, from such limitations. To this end, the use of hardware to accelerate the cryptographic operations can prevent the network from computational overhead.
- *Performance*: Due to the reliance on real-time data shared by most of the IoT ecosystems, such as smart grids, smart cities, smart homes, *Wireless Sensors Networks* (WSN) etc., the system's performance can be considered as significant as its security. To secure the future of efficient IoT ecosystems against failures, it is imperative these systems to be self-managed and self-regulated. However, a possible increase in the system's performance should not result in lowering its security. Moreover, the number of IoT devices is constantly increased and this will result in the generation of more IoT data, estimated to be 79.4 zettabytes until 2025. Therefore, it is compulsory that the corresponding IoT ecosystems will be able to accommodate any possible expansion of their network and handle an abundant number of messages with high performance metrics.

The current threat spectrum necessitates the use of multiple sophisticated security mechanisms for the IoT ecosystem. Various researchers envisage the blockchain technology as the silver bullet to strengthen the security of the IoT. Therefore, it is important to get familiarised with this tech-nology, before proceeding further.

## 2.2. Blockchain technology characteristics

The blockchain is an innovative technology, which has proven to provide considerable advancements in several areas by inherently providing interoperability, decentralisation, security, auditability, privacy, persistence and sustainability; it can be considered as an autonomous root-of-trust, to transfer information from one entity to another, even in adversarial environments. For this reason, the blockchain is treated as a highly secure technology to share information and execute transactions by forging not only a distributed sequence of records but also a linear and chronological order of blocks.

The blockchain technology basically realises a distributed service without depending on any central authority to enforce trust between the participating entities. It is meant to provide an immutable service for preserving records between a set of mutually distrustful entities. The security properties that are fulfilled by the blockchain technology are unprecedented really enthusiastic. The pioneering work of Nakamoto [165] has tremendously affected the digital payments around the world. Therefore, due to the unique and desirable properties of blockchain, this technology along with the applications that are built on top of it, are envisioned to revolutionise not only the majority of the financial services but also various other areas.

From its infancy, the blockchain technology empowered segments of executable code to define operations that have been acknowledged by all the participating entities, i.e. the smart contracts. The smart contracts are computer programs which are embedded in the blockchain to digitally establish, verify and enforce that a contractual agreement is executed on a given condition. Therefore, each node can run the contract locally, propose the result to the network and then, cooperate with it to decide which result is going to be included into the blockchain [84]. Smart contracts however can provide features only as far as the blockchain technology can, and test the limits of our modern lives beyond crypto-currency. Therefore, as a revolutionised technology of our today's society, the smart contracts can be adopted in many business areas, such as the supply chain, the intellectual property, the digital asset exchange and the IoT ecosystem.

Blockchain networks are categorised into different settings. In contrast to permissionless, in which each node can be part of the network (Bitcoin, Ethereum etc.), the permissioned settings can be either public or private and are defined by the fact that the participating entities, who regulate and amend the blockchain, are identified and can be held accountable for their actions. The key insight behind this concept is to fulfil the necessary flavors of privacy and other requirements of the blockchain applications, as well as to provide beneficial access control mechanisms along with some means of identification.

Between the various components that are included in a distributed blockchain network, the consensus protocol is the key process that facilitates the network's security and most importantly ensures that all the participating entities achieve atomic broadcast, i.e agree upon the total order of the transactions, without any trusted authority. Therefore, various design choices of the protocol can significantly impact the network's security (i.e., the fault tolerance) and its performance (i.e., the transactions' throughput, latency and the system's scalability).

## 2.3. Related work

To date, several surveys [46, 54, 62, 63, 93, 110, 114, 123, 131, 152, 159, 164, 191, 196] have been conducted regarding the blockchain technology and its integration with the IoT ecosystem. However, there is no promising work that evaluates, in a quantitative manner, this integration. Several of the aforementioned surveys are focused on general applications of the blockchain technology or discuss various technical aspects that concern the digital crypto-currencies. Some of these surveys are focused only on the consensus' security and provide qualitative information - instead of quantitative - about the performance metrics, without taking into account other important aspects, such as the smart contracts, the platform's security and the privacy preservation mechanisms that have been implemented.

The authors in [46] provide an intuition into the use of smart contracts for the IoT. The paper focuses on the use cases of a blockchain-enabled IoT ecosystem and discusses the requirements of this integration. These requirements include the low transactions' throughput and the high block confirmation time in consensus protocol, as well as privacy and legal matters that are correlated with the smart contracts. The authors in [54] proposed a lightweight model for the integration of the blockchain technology to a smart home, focusing only on the limitations of the proof-of work consensus protocol and presenting solutions to avert issues that may occur from the computation overhead and the block confirmation time, as well as scalability issues.

In [63] and [159], the authors conducted a comprehensive review of how to integrate the blockchain technology to the IoT. However, these papers discuss only the challenges, the recommendations and all the possible optimisations that can affect the architecture and the development of this integration. In [62], the authors examined the requirements of the IoT and proposed a framework that identifies suitable blockchain solutions for different IoT applications.

Khan et. al., [93] conducted a survey regarding the security issues that might arise in an IoT ecosystem and outlined only how these issues can be addressed by the blockchain technology. Lao at. al. [110] surveyed a number of consensus protocols and blockchain platforms for the IoT ecosystem and compared them in terms of the given transaction's throughput and the consensus security. The paper focuses only on these two factors in order to evaluate whether the blockchain technology can be implemented in the IoT. In a recent paper [114], the authors partially examined the security and privacy risks of the most prominent blockchains in all the layers of the blockchain architecture. This survey paves the way to further consider more aspects regarding the

**Table 1**
Related work regarding the surveyed factors for the adoption of the blockchain technology to the IoT ecosystem

| References | Challenges in the IoT | Consensus' security | Platform's security | Smart contracts' security | Performance metrics | Privacy mechanisms | IoT suitability |
|---|---|---|---|---|---|---|---|
| [46] | ✓✓ | ✓✓ | – | ✓✓ | – | – | – |
| [54] | ✓✓ | – | – | – | – | – | – |
| [62] | ✓✓ | ✓ | – | – | ✓ | – | / |
| [63] | ✓✓ | ✓✓ | – | – | – | ✓ | – |
| [93] | ✓✓ | – | – | – | – | – | – |
| [110] | ✓✓ | ✓✓ | – | – | ✓ | – | – |
| [114] | – | ✓ | ✓ | ✓✓ | – | ✓ | – |
| [123] | ✓✓ | ✓✓ | ✓ | – | ✓ | – | / |
| [152] | ✓✓ | ✓ | – | – | ✓ | – | / |
| [159] | ✓✓ | ✓ | – | – | – | – | – |
| [164] | ✓✓ | ✓✓ | – | – | × | – | / |
| [191] | ✓✓ | ✓✓ | – | – | ✓✓ | – | – |
| [196] | ✓✓ | ✓✓ | ✓ | ✓ | ✓✓ | – | – |
| Our work | ✓✓ | ✓✓ | ✓✓ | ✓✓ | ✓✓ | ✓✓ | ✓✓ |

✓✓ denotes that the factor is fully examined
✓ denotes that the the factor is surveyed, but not fully examined;
– denotes that the factor is not surveyed;
× denotes that the performance metrics are given in a qualitative way;
/ denotes that the IoT suitability is undefined security issues that might arise in a blockchain implementation. Perez et. al., [152] proposed an overview of a small number of consensus algorithms and examined their suitability to an IoT ecosystem.

In another work, [123], the authors compared a number of consensus protocols and few blockchain platforms and, based on the consensus' security and the performance factors, made an attempt to define a suitable technology for the IoT. However, this paper does not focus on the privacy issues, the platform's security or the smart contracts. In [131], the prerequisites and the issues that might occur when the blockchain technology is implemented in a smart home ecosystem is being discussed, highlighting the design practices that eventually will lead to better integration of a smart home ecosystem with the blockchain technology.

In yet another work [191], a survey on consensus protocols and blockchain platforms for IoT applications is given. The authors highlighted the advantages, the drawbacks and the key challenges of the blockchain technology in IoT applications.

In a similar work to ours [196], a taxonomy of the most prominent consensus protocols for the IoT was proposed. The comparison is based on the performance and adversarial tolerance of the surveyed consensus protocols, as well as, on the platforms' type. However, several other factors for IoT suitability are not mentioned in this paper, such as the privacy preservation mechanisms that have been implemented in each blockchain platform and the security of the smart contracts.

To the best of our knowledge, the only works closest to ours are [123, 164, 191, 196]. In a recent survey [164], Salimitari et. al., examined a number of consensus protocols and blockchain platforms for evaluating their suitability to the IoT ecosystem. However, the corresponding suitability of each consensus protocol for the needs of the IoT seems to be undefined and without strictly defined criteria. The performance metrics and all the critical factors are presented in a qualitative manner without considering other important aspects for the blockchain technology integration, such as the platform's security, the security of the smart contracts and the privacy preservation mechanisms.

Therefore, to cover the gaps in the literature regarding the blockchain's integration in the IoT and particularly in a smart home ecosystem, there is a need to carry out a comprehensive survey to find out how existing blockchain solutions integrate with the IoT ecosystem and how suitable these solutions are in a quantitative manner. For this reason, a methodical review for the available blockchain solutions to determine their ability to meet the requirements and address the challenges of the smart home ecosystem, based on certain efficiency, security and privacy factors, and an evaluation of their suitability for our reference IoT ecosystem is necessary. Therefore, there are many factors that distinguish our work from previous relative work and they are presented in Table 1.

## 3. Blockchain architectures for smart homes

Smart home, the most popular and promising use case of the IoT, has already become a significant part of the aforementioned digital revolution. This smart residence encompasses a network of appliances and devices that belong to various application areas, such as lighting control, appliance control, entertainment, assisted living, climate control and safety systems [112]. The combination of these appliances to a particular use case can aid to create innovative, automated, and smart services for residents through distributed and collaborative operations. However, these smart devices pose potential risks to privacy, since the IoT data involved in smart homes is mostly personal. In this context, many research studies have been demonstrated that most smart home appliances lack basic security features and can be easily compromised [126, 131]. Moreover, most current smart homes, depend upon centralised cloud services with a single access point for data storing, processing and management. Such architecture creates data transparency, trust and privacy-related issues, since the home residents have no control over their own data [112, 172].

The emergence of blockchain as a secure and transparent structure for protecting data has inspired fresh thinking about the benefits of applying this technology to the area of the smart home. The functionalities provided by the blockchain technology can significantly *a*) enhance the smart home's security, *b*) satisfy the necessary flavors of the home-owner's privacy and *c*) resolve integrity issues [147]. In addition, a blockchain-enabled smart home ecosystem can significantly improve the security level of this domain by providing trusted, transparent, and secure sharing services and therefore, assure the authenticity and traceability of the personal data exchanged amongst the IoT devices. Another benefit that is brought by this integration is the development of a trusted and distributed authentication network, in which the source of data can be uniquely identified at any time and remain immutable [159]. All these benefits, amongst many others, demonstrate that the blockchain technology can face the security and privacy requirements in a smart home ecosystem (Figure 1).

When the blockchain technology is implemented in a smart home, three possible scenarios can occur [159]. In the first scenario, the interactions between the IoT devices are stored on an external database and only a part of the associated data is stored on-chain. Whereas, in the second scenarios, all the interactions are stored on the blockchain along with the associated data, keeping an immutable record of all the interactions that occur in the network. The third scenario is a hybrid design, in which only a part of the IoT data and the corresponding interactions are recorded.

The first scenario is the optimum case in terms of latency, due to its off-chain implementation. However, it can only be applied in cases where the IoT data is reliable and the interactions between devices and the appliances occur almost instantaneously [159]. The second scenario is more applicable, but in this case all the interactions are stored on the blockchain. In this case, bandwidth and performance issues can easily arise. The last case is gaining more attention from academia and industry, but choosing which interactions should be recorded is considered to be another challenging task. In this article, we present the most relevant implementations that have been developed by individual researchers and/or in the context of research projects.

### 3.1. Cyber-Trust blockchain

Cyber-Trust blockchain (CTB) is a private blockchain platform, built on-top of Hyperledger Fabric and designed to facilitate a secure collection and preservation mechanism of forensic evidence for a smart home ecosystem [38]. The system implements a private forensic evidence database, where the captured evidence is stored, so that it can be used in a court of law. The blockchain platform stores evidences' metadata, which are critical for providing the important services; and interacts via smart contracts with the different entities involved in an investigation process, including law enforcement agencies, internet service providers and prosecutors. The proposed blockchain platform has been developed in the context of the Cyber-Trust project [50] that aims to develop an advanced detection and mitigation platform for trusted smart home ecosystem. The main idea of this mechanism is that if an IoT device is compromised, an *Intrusion Detection System* (IDS) is initiated to collect relevant information that can be used as digital forensic evidence during the investigation process. The IDS can operate not only at the device's, but also at the network's level to facilitate the collection of evidence from various independent sources. During pilot tests in this blockchain-enabled SOHO ecosystem [38], it was found that when each SOHO is under attack, it creates 50 alerts per second, with each alert to be on average 1.5 kbytes. Having $10^4$ SOHOs, it was found that more than 2000 TPS were required for a sustainable implementation. This mechanism also follows a combination of the first and the third integration scenarios, in which only the critical information that has been captured by the detection system is recorded off-chain and the metadata are recorded on-chain.

### 3.2. IoTex blockchain

IoTex [183] is a decentralised blockchain solution that aims to empower the IoT ecosystem by an end-to-end trust, where all IoT devices, humans, machines and businesses can securely exchange data at a global scale. This system provides a well-balanced distributed network that addresses both privacy and scalability issues. A fast consensus protocol with low-cost transactions is combined with a lightweight cryptography mechanism to conserve energy, storage and computation resources. The main idea of this project [183] is analogous to that of the Internet, but in this case, the IoTex blockchain locally supervise a specific group of IoT nodes, which are part of the smart home ecosystem. This implementation can provide various benefits to the network, but the main goal is to safeguard the home-owner's privacy.

### 3.3. GHOST blockchain

The GHOST protocol [102] aims to deploy an efficient security framework by implementing a private Ethereum network in a smart home ecosystem. The blockchain implementation follows the second integration scenario, which enables each IoT device to act as a regular blockchain peer and as a smart home gateway middle-ware, at the same time. The computational limitations are addressed by assigning in each node a different work to perform; therefore, the nodes with limited computational capabilities act as lightweight nodes and only devices with adequate computation power (i.e. the full nodes) participate in the consensus process, thus making the GHOST protocol a functional implementation.

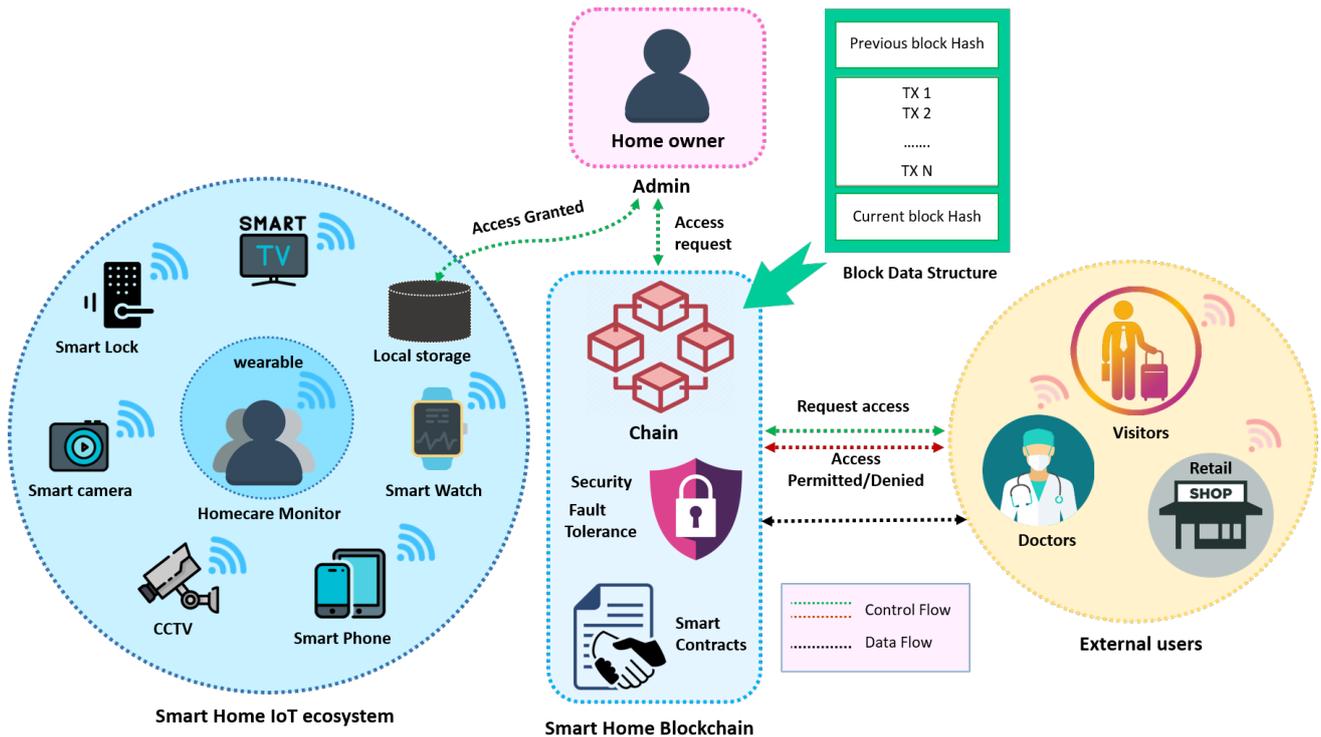

**Figure 1:** Blockchain integration to the smart home ecosystem.

### 3.4. Atonomi blockchain

The key idea behind the Atonomi project [14] is to provide an innovative embedded solution to secure IoT devices by rooting their identity and their reputation on an immutable distributed ledger. This is accomplished by creating and deploying a network that maintains decentralised consensus for the transactions of the IoT devices on the Atonomi network. The Atonomi architecture, which is built on-top of the Ethereum platform, is fully extensible by programmers across the IoT verticals and capable to safeguard the vast realm of the IoT, ranging from the smart home ecosystem to the wide-industrial IoT landscape. In the context of the smart home, the home-owners can register their IoT devices to the Atonomi network and secure them from various attacks, but also from other devices that might have been compromised.

Various other individual researchers have also introduced intriguing blockchain-based smart home applications [6, 18, 54, 86, 119, 138, 195]. In this context, a novel instantiation of blockchains for smart homes is proposed in [54]. The main goal of this work is to preserve home-owner's security and safeguard its privacy by using a local and private blockchain that provide secure access control to the devices and to the user's IoT data. The proposed architecture involves three core tiers, namely: the overlay network, the smart home, and the cloud storage layer. In this solution, a highly resource device centrally manages all the communication that occurs in its smart home ecosystem, either internal or external. The shared overlay network has been proposed to inhibit multiple smart home-owners in a single blockchain network and reduce their operating costs. For these challenges to be achieved, the nodes are gathered into clusters, with each cluster to appoint a Cluster Head to be the leader. This architecture follows the second integration scenario, in which a local blockchain mechanism is used to keep track of all the transactions that occur in the smart home, along with the local and the cloud storage solutions that store and share the IoT data.

## 4. Requirements for a blockchain-enabled smart home ecosystem

This section studies the major issues needed to be addressed for the integration of the blockchain technology in a smart home. This integration is not trivial, since blockchains have been designed for a use case scenario with capable computers, which is far from the smart home reality [159]. Currently, various commercial projects provide to the smart home-owners the ability to purchase their own blockchain-based solutions. However, some of these solutions lack in terms of security, performance and privacy protection, with no explicit instructions to be included for their integration. The primary challenges concern the implementation of blockchain applications that better suit to these requirements. Therefore, the feasibility of distributed applications in smart homes, needs to be fully realised under real world deployments. For these reasons, we explore and cate-gorise the prerequisites that are needed to be taken into account before the blockchain technology is integrated in a smart home ecosystem. Our methodology is illustrated in figure 2, in which the requirements taken into account - and the sections are presented in order to evaluate the suitability of the blockchain technology to a smart home ecosystem.

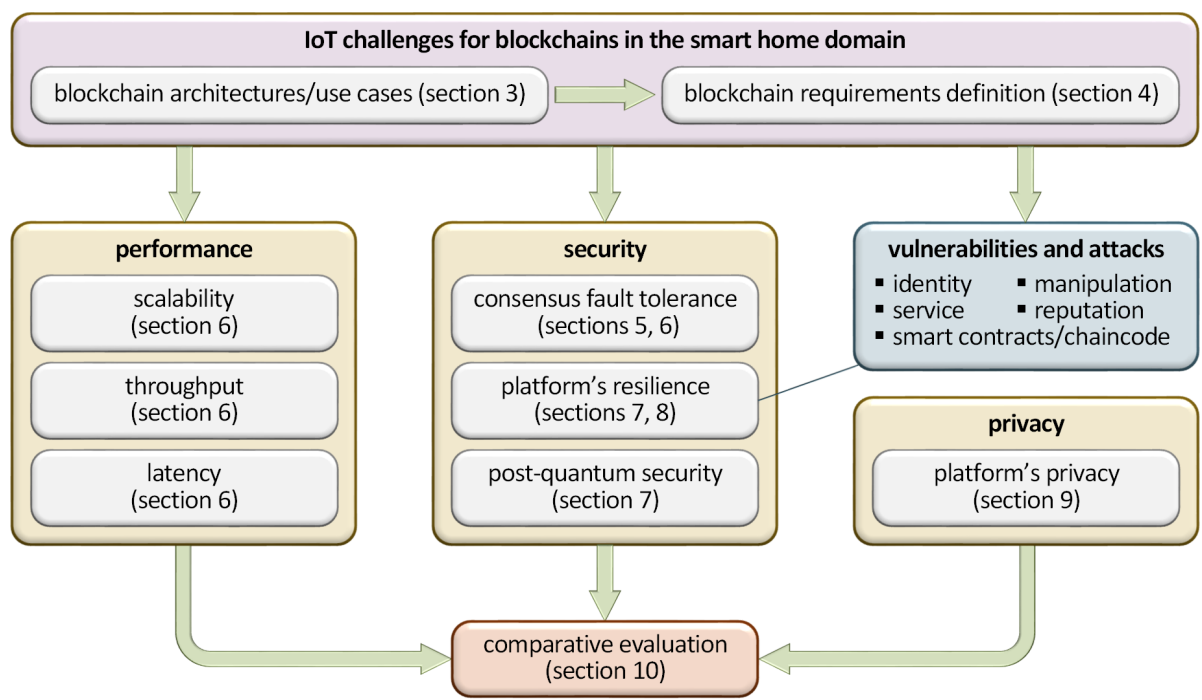

**Figure 2:** Blockchain requirements and paper's methodology.

### 4.1. Performance requirements

In a smart home ecosystem, the IoT devices can be categorised as data generators that create data traffic, and data consumers that respond to a generated event. For instance, a smart thermometer regulates the temperature to create or maintain a comfortable environment for the home-owner by disseminating instructions to the smart air-conditioning devices. The implementation of a blockchain mechanism in this case, might hamper responsiveness by its performance with the generated IoT data to be consumed at a later time than the expected. If the confirmation time of blockchain transactions that record IoT operations is stalled, significant delays might occur in the future control and communication with the IoT devices. Therefore, the IoT data should be immediately stored on a high performance shared ledger without impeding the network's efficiency.

These requirements, in terms of performance, are studied in many research papers [8, 9, 21, 27, 31, 32, 57, 65, 66, 110, 122, 134, 139, 145, 160, 174, 176, 184, 186, 188, 189, 197, 200], and are categorised as: 1) Scalability, which is the ability of the network to support an amount of users and any possible increase of its size, without degrading its performance. 2) Transactions' throughput, which defines how many *transactions per second* (TPS) can be processed during the consensus process, and 3) Latency or *block confirmation time* (BCT), which defines how much time is required for a transaction to be confirmed and included to the distributed ledger.

In a smart home ecosystem, the IoT devices might need to join or leave the network on-the-fly for various application scenarios. Scalability is often defined as a broader concept of the network's performance and it is the qualification of the blockchain network to maintain throughput when the consensus nodes are increased. For instance, some protocols can only work well in a limited and fixed-size network, in which the number of the network's nodes cannot be more than a few tens and easily modified. Therefore, deploying a huge blockchain network in a smart home, without the nodes being able to leave or join the network, can lead to its performance degradation and compromise the network's security.

Apart from scalability, the transactions' throughput and latency are some of the most critical factors to be considered in a blockchain-enabled smart home deployment, since it is required that a transaction is being broadcast and appended into a block within seconds or even less [131, 164]. Therefore, traditional implementations (such as, Bitcoin's) and consensus protocols (such as the Nakamoto's consensus) might not be applicable for delay-delegate smart home applications, in which small networks delays can cause severe consequences.

In our work, we evaluate the transactions' throughput as *high*, if more than 2000 TPS can be processed, and the transactions' latency also as *high*, if the time that is required for each transaction to be confirmed is more than 10sec. The thresholds derivation occurred based on the researched blockchain smart home ecosystems and the expected volume of events/transactions being generated in a typical smart home ecosystem, according to the approaches presented in section 3. For example, in [12] and [17], blockchain-enabled smart home ecosystems were proposed based on the Ethereum platform. In both of these implementations, the authors mentioned that their systems may undergo a challenge in time-delegate conditions, since it takes enough time for each transaction to be confirmed [12, 17]. Moniruzzaman et. al., [131] even claims that the BCT of 3.9sec needs to be further reduced for a sustainable blockchain-enabled smart home ecosystem. Therefore, BCT greater than 10sec might not be so suitable for time-sensitive applications. On the other hand, the required transactions' efficiency is rather based on the integration scenarios. For instance, much higher throughput is required when all the interactions that occur in the smart home are stored on the blockchain than in a scenario, in which only the critical alerts are being stored on the blockchain. However, in our work, different values have been given for the different performance metrics for both the transactions' throughput and confirmation time. An acceptable example, in terms of performance, is proposed in [4], in which the achieved throughput is greater than 2000 TPS and the BCT is less than 1sec.

### 4.2. Storage requirements

A smart home includes always connected IoT devices that communicate and target services. These devices do not require any tremendous computation power and storage area. Therefore, their limited/non-existing memory or storage capacity make the preservation of the IoT data a significant issue. The integration of a huge processing mechanism, such as a blockchain platform, into a resource-constraint network might not be seen as a suitable solution. However there are ways, in which these limitations could be alleviated or even avoided.

Currently, in a smart home ecosystem, a huge amount of IoT data is generated, whereas only a limited part of it is useful to extract knowledge and to generate actions. Furthermore, several techniques have been suggested to filter and compress the data derived from the IoT devices and address the storage limitations. Therefore, savings in the amount of the stored data of the smart home can be achieved. Another possible solution to this issue is the separation of the actual data from the metadata. Having the data stored in a secure database and the hashes of this data on the blockchain, the storage issues can be easily solved, whilst the system's security is being preserved [38]. Throughout the paper, we consider storage requirements to be strongly linked to the maximum expected amount of transactions generated by a blockchain platform; thus, TPS also serves to measure the storage needs for a typical transaction size of $\approx$ 1KB (we are *only* interested in the *order of magnitude* here), and this is eventually reflected in a platform's scalability; (*see* section 10 for the details).

### 4.3. Privacy requirements

The right to privacy has been recognised as a fundamental human right by the United Nations Declaration of Human Rights, the International Convenant on Civil and Political Rights, the Charter of Fundamental Rights in European Union and other international treaties. Privacy is strongly related with the personal data protection; as stated in the Charter of Fundamental Rights, personal data (i.e. data relating to an identified or identifiable natural person) must be processed fairly for specified purposes and on the basis of the consent of the person concerned or some other legitimate basis laid down by law under Article 8(2) of the Charter. The Charter furthermore provides that everyone has a right to access personal data relating to them, including a right to have such data rectified or deleted [149].

The *General Data Protection Regulation* (GDPR), being applicable since May 2018, constitutes the main legal instrument for personal data protection in Europe, establishing a detailed framework that harmonises data protection across the European Union. Most importantly, GDPR also applies to the processing of personal data of individuals being in the European Union by organisations not established in the Union, in case that the processing activities are related to either the offering of goods or services or the monitoring of their behaviour (as far as their behaviour takes place within the European Union). The GDPR codifies the basic principles that need to be present when personal data are being collected or further processed, in order to promote the fundamental right of personal data protection.

In a smart home ecosystem, many IoT devices process confidential personal information, which poses risks for the rights and freedoms of individuals; for instance, if an IoT device is linked to a smart home-owner and is related with health data, such as in an e-health scenario [55], then an improper use of this information may harm the individual even if the device is not attacked. Therefore, it is essential to ensure each user's essential right to privacy. In general, the necessary flavors of privacy cannot be easily guaranteed when data derived from IoT devices are being processed and, thus, a careful design approach should be followed [70]. Such privacy issues entail more effort, since they start from the data collection process and extend to the data preservation mechanisms.

The question of compatibility between blockchains (not only in IoT applications but in any case) and GDPR is not always an easy answer. For example, blockchains are presented as the ideal solution to counter the identity management issue in a smart home ecosystem [54] or when the home-owners want to allow, refuse and withdraw access to their personal data according to different cases of potential use [136]. However, in general, the inherent properties of blockchains may also give rise to privacy threats. For instance, the transaction's linkability is a privacy threat, in which the data (transactions) that are stored on the blockchain can be linked and allow a reconstruction of the user's profile (which in turn can be used for other purposes, without the concerned user being aware of this). This is an obvious privacy threat in permissionless and public settings, where the entire blockchain is visible to an adversary. As stated in [149], the aforementioned compatibility issue can only ever be determined on a case-by-case basis that accounts for the respective technical and contextual factors (such as the governance framework).

Therefore, for securing the sensitive IoT data from unauthorised entities, it is required the implementation of permissioned and private settings. Moreover, to tackle with other risks of malicious access to the users data (which is both a privacy as well as a security issue), appropriate cryptographic primitives may be adopted by the resource constrained IoT devices. In addition, depending on the specific data protection requirements induced by the context of the application, several privacy enhancing technologies relying on advanced cryptographic techniques can be implemented (i.e in cases where the data in a blockchain corresponding to a specific individual need to be indistinguishable from the relevant stored data corresponding to other individuals, in other to mitigate the risk of illegitimate profiling). Such technologies are being subsequently discussed in section 9, in relation with the relevant blockchain platforms that support them.

In our work, privacy is mapped as *medium*, and *high* in order to satisfy the user's privacy requirements in a blockchain - enabled smart home ecosystem. Privacy is defined, based on the nature of the blockchain platforms: permissionless; denoted in the text as "PL" and permissioned, which is denoted as "P". Towards this categorisation, privacy preservation mechanisms have been implemented in various blockchain platforms, even in Bitcoin, to serve various purposes. Nevertheless, permissioned blockchain platforms are more capable to satisfy the necessary flavors of privacy in this particular ecosystem and therefore, privacy is mapped in this way.

Last but not least, a smart home owner may exercise its data rights on the processing of its personal data - including the case that the personal data are stored on a blockchain. Due to the immutability that the blockchain technology provides, the information that it is stored on a distributed ledger cannot be easily erased. This inherent property of blockchains is another privacy issue related with the so-called "the right to be forgotten" in the GDPR. Therefore, privacy laws that concern the user's privacy with respect to the right of deletion of data constitute another significant challenge to be tackled in an smart home environment and will be an even greater if they are used in integration with the blockchain technology.

### 4.4. Security requirements

The increasing number of IoT devices comes with new cyber attacks that have serious effects, thus necessitating the development of smart home ecosystems with more sophisticated security. The IoT devices interact with each other through a gateway using various wireless communication protocols, which usually paves the way for cyber-attackers to perform eavesdropping and several other attacks. The IoT home devices (such as web-cameras, smartphones, etc.) may be compromised and provide unauthorised access or allow the attacker to collect sensitive information. The key security issues in smart home ecosystem are related with weaknesses in authentication and authorisation mechanisms, vulnerabilities in the access control system and configuration issues. The traditional smart home ecosystems systems are mostly centralised and linked with a cloud server; hence if the central server is compromised, the overall security collapses. Therefore, to overcome these issues, various solutions have been proposed that include additional security layers in existing smart home ecosystems [90].

Blockchain can be seen as the key technology to provide the fundamental security enhancements to secure the smart home ecosystem and diminish the security concerns regarding the authorisation and authentication, integrity and confidentiality of IoT data, as well as, the single point of failure. However, an essential challenge for the integration of the blockchain technology to the smart home ecosystem is the reliability and the trustworthiness of the IoT data [99, 169]. Due to the immutability of blockchains, if the IoT data that are being stored to the blockchain are already corrupted, then they will remain corrupted. Such situations can arise not only from malicious behaviour, but also from several other possible failures related with the environment or the participants. Therefore, the IoT devices should be exhaustively tested before this integration and appropriate security techniques should be adopted to detect possible device failures.

Despite the fact that most of the blockchain platforms can resist traditional attacks in an optimum way, the adversaries create each day new ways to mount successful attacks [100]. Various blockchain platforms and consensus protocols have been thoroughly analysed in terms of security and several vulnerabilities and security issues have been discovered. One of the most common attacks in distributed systems is the double spending, in which the adversary manipulates the consensus protocol used and attempts to insert malicious transactions to the shared ledger. Therefore, 1) the consensus protocol that is going to be implemented in a blockchain-enabled smart home ecosystem should be able to address any possible malicious behaviour; 2) the blockchain platform should not be susceptible to cyber-security attacks that aim at breaking its integrity and availability, and 3) the smart contracts that execute security operations (such as authentication and access control) should be secure, in such a way that there will be no backdoors for an adversary to compromise the network.

In this paper, we evaluate the consensus' and the platform's security by mapping them to a three-valued scale, namely *low*, *medium*, and *high*. For the consensus protocols that can tolerate crash failures and malicious behaviour (resp. only crashes) security is considered to be high (resp. medium), whereas for the centralised consensus protocols that are destined for developing testing and therefore, the security mechanisms that they possess are considered to be low. On the other hand, each platform's security is based on its capability to address some of the most common attacks that can occur in a blockchain-enabled smart home ecosystem.

## 5. Fault tolerance in consensus protocols

Distributed systems are usually disciplined by a set of clients and services, where each service utilises one or more servers to extract information or execute operations that are requested from the clients. Using a central server is the easiest way to fulfil the necessary needs and the implementation of a service, but it posses a major concern about the security of a smart home ecosystem, due to the fact that it becomes a *single point of failure* (SPoF). Thus, to avoid centralised faults, multiple servers should be deployed to implement a Fault Tolerant service with *state machine replication* (SMR) [166].

The pioneering work of Lamport [108], who introduced the Byzantine Agreement, triggered the research of developing algorithms in order to exploit and construct resilient distributed systems. Nonetheless, several blockchain systems deviate from the classical SMR in crucial ways. Many distributed applications run simultaneously and can be deployed at any time, even if the embedded application code is untrusted or occasionally malicious. The key idea to provide security in blockchains is to reach agreement on a single request from a client, which is the core functionality of a consensus protocol. In the context of blockchains, a consensus protocol or as it is commonly known as an "atomic broadcast", provides a total order of the disseminated messages and propagates them to the network's peers/nodes.

### 5.1. Crash fault tolerance

An atomic broadcast certifies that all the legitimate peers output or deliver the identical array of messages by means of the deliver event. Accurately, considering a set of $n$ peers in the network, it certifies that the properties validity, integrity and total order are fulfilled [76].

The way to achieve consensus (i.e. to realise atomic broadcast) in distributed systems that are vulnerable to $t < n/2$ node crashes is to adopt consensus protocols known as the *viewstamped replication* (VSR) [142] and Paxos [107] family of protocols. To provide security, this family is characterised by the same rules. In each round a leader is elected or voted to create a new block and if the ongoing leader crashes or even if the peers in the network suspect that the leader has crashed, the leader is reinstated by proceeding to the next round. This family of protocols is known today as *crash fault tolerant* (CFT) consensus protocols and it guaranties that a set of failing nodes $t < n/2$ does not impact the system.

### 5.2. Byzantine fault tolerance

Consensus protocols with the purpose of tolerating byzantine nodes, which are subverted by a malicious actor and avert the common goal of reaching agreement, have recently emerged. In the *Byzantine fault tolerant* (BFT) consensus protocols family, the most common protocol is the *Practical Byzantine Fault Tolerant* PBFT [41], which can be displayed as a blossom of the VSR/Paxos [107, 142]. In a network comprised of a set of $n$ nodes, the PBFT consensus protocol [41] can tolerate $f < n/3$ subverted nodes using a progression of rounds with a unique leader within each round. Under the assumption that the BFT protocols seem to be more secure than CFT, vast research work has focused on the improvement of the PBFT with BFT-SMaRt [28] to be considered one of the most advanced and scalable BFT consensus protocols.

## 6. Consensus protocols in blockchains

The consensus protocols are mechanisms that define a set of rules to provide truthful services among participants without mutual trust to collectively maintain a synchronised state. Due to its nature, a consensus protocol provides incentives for the participating nodes to be honest and create or add new blocks to the blockchain. Such consensus mechanisms are described in this section and are mostly used in cryptocurrency-based applications, where the need to behave honest and not deviate from the predefined process might cost significant losses. Moreover, some of these blockchain solutions provide a desirable, suitable and high performance network required by the smart home ecosystem only if, the monetary reward that incentivises the consensus nodes to behave honestly is replaced by IoT-based criteria.

### 6.1. Proof of work (PoW) and its variants

The Nakamoto's consensus [165] is the key technology that revolutionised the digital currency exchange, since it is implemented on several other cryptocurrency-based ledgers, such as Litecoin [116], Monero [141], Ethereum [194], etc,. Several other hybrid variations of the protocol also exist, in which the peers use their computational power to elect a leader, e.g. Bitcoin-NG [58] and Elastico [121]. Nakamoto's consensus forces each node to calculate a huge amount of hashes and mine a block only if the node finds an appropriate value that satisfies a specific property which is difficult to be achieved. This mathematical problem serves as a means to establish consensus and verify the validity of blocks on the network before accepting it. Moreover, the validation of blocks necessitates the assistance of other principles as well, such as 1) the longest chain rule, where only one chain is commonly adopted by network, and 2) the incentive rule, where only a single node will gain the reward by finding the next block in the chain. PoW can tolerate at most 50% of its network's total computational power to be used for malicious purposes towards solving the aforementioned puzzle. Therefore, each node is required to spent a vast amount of energy to adhere to the protocol. Therefore, the spent computational power is the necessary regulation for each node to adhere to the protocol, by building trust between all the participating nodes.

As to the terms of performance, even in the same consensus protocol, slightly changing some parameters might result in different throughput and BCT. Bitcoin can provide 7 TPS processing rate, with each transaction to be included in a block in 10min. Similarly, Litecoin can achieve 56 TPS and BCT of 2.5min, while Ethereum and Monero seem to be better solutions with 25 and 1700 TPS, as well as, BCT of $10 - 15$sec and 2min, respectively. However, several other variations of the Nakamoto's consensus have been proposed to reduce the computational requirements and provide increased performance. Bitcoin-NG [59] can achieve 200 TPS and BCT of 10min, while Elastico [121] adds to its distributed ledger 16 blocks every 110sec.

### 6.2. Proof of stake (PoS) and its variants

To address the limitations with respect to the computational power that the Nakamoto's consensus introduces, a different type of consensus has been proposed. The key idea of PoS is the concept of "stake", where the nodes that participate in the consensus process, lock into an escrow account, which is a specific amount of coins. The concept of the stake serves as a guarantee that each node will behave in accordance with the protocol rules and will not deviate from it. Therefore, the users that possess more stake are more incentivized - with stake's loss - to safeguard the system's reliability, since it is less possible to be turned and become malicious nodes. PoS can tolerate at most 50% of its network's total stake to be used for malicious purposes. Despite of this, the classical PoS protocols cannot achieve high performance metrics; Snow-White [25] and Ouroboros [94] are quite popular protocols, but in terms of performance Snow-White achieves at average 125 TPS and BCT of 10min, Ouroboros can reach 256 TPS, with each transaction being confirmed in 2min. To counter the low performance of PoS protocols, several other variations have been proposed with Algorand [69] to be built on top of byzantine agreement. Algorand, a platform that aims to solve the "decentralisation, scalability, and security trilemma", attaches a cryptographic proof to each new block and checks the eligibility of the block proposed to be chosen, while this probability is directly proportional to its stake. In terms of security and performance, Algorand can tolerate malicious behaviour and provide high scalability with more than 1K TPS and each transaction to be confirmed in 5sec.

### 6.3. Delegated proof-of-stake (DPoS)

Implemented in Bitshares [32], EOSIO [57], Lisk [115], Tezos [184] and Tron [186], DPoS [111] is a reputation-based consensus protocol, considered to be a more efficient and democratic version of PoS. In a similar way, the protocol proceeds in rounds and within each round, a leader is selected (voted) according to the blockchain platform that the protocol is being implemented to. Although it derives from PoS, there are many differences that make DPoS exceptional. The network's users vote with reputation scores to choose a class of delegates that will be authorised to participate in the consensus process. Amongst a set of delegates, a leader is elected and incentivized to adhere to the protocol and create honest blocks. With the creation of a new block, the leader is either rewarded, if the block is honest, or penalised and blacklisted in any other case. The protocol can withstand 33% of the delegates to be malicious. With the ambition to be elected, the delegates can promise various levels of rewards to the token holders. A token holder can entrust a delegate with its own stake, by casting a vote via a blockchain transaction. Then, the delegate acquires the stake power from all of its voters and behaves as their fronted in the consensus process. As mentioned, DPoS' performance metrics differ across the blockchain platforms that it is implemented, but the scalability of the protocol is similar to PBFT's. Bitshares achieves the tremendous transactions' throughput of 100 thousand TPS with each transaction to be confirmed in 1sec, EOSIO's throughput is measured to be $1-6$ thousand TPS and its BCT is less than 1sec. DPoS has also achieved high performance metrics in Tron with throughput greater than 2 thousand TPS and BCT $1-3$sec. Lisk and Tezos are the exception of the DPoS implementation and can achieve low throughput of 2.5 and 40 TPS, while the BCT is measured to be 6 and 30min, respectively.

### 6.4. Kafka protocol

Kafka [103] is a distributed publish/subscribe messaging pattern that can transfer large amount of log data with significantly low latency. Kafka's key components are the producers, the topics, the consumers and the brokers. The recorded information is published from the producers to a stream of messages called the "topic", which is a partition of segments of files. The messages are stored from the brokers to the latest segment file and when the producers publish messages to the partitioned logs, only the subscribed consumers can consume these messages, sequentially, by simply making requests to the brokers. At high level, the Kafka's conceptual configuration is based to the leader – follower setting, in which the transactions are replicated from the elected leader to its followers and if the leader crashes or even if there are suspicions that the leader has crashed, then one of its followers takes command. Kafka is a CFT consensus protocol that can address less than 50% of network's crashes and it is mostly known for its implementation in Fabric [9] and FastFabric [72], in which it has been shown to achieve throughput of 3.5 and 20K TPS, respectively, with each transaction being confirmed in less than 1sec.

### 6.5. Raft protocol

The Raft consensus protocol [143] is currently implemented in Fabric [9], Quorum [153] and Corda [155]. Acquiring its security properties from Paxos [109], it is a CFT protocol based on the "leader-election" model. Raft establishes consensus by electing a leading node to obtain the incoming entries from the clients and replicate them. The protocol is separated into three phases allowing a strong and coherent leader election process. These phases are the leader election, the log replication and the safety [143]. The time in Raft proceeds in arbitrary time periods, called "terms", with each term to be defined by an increasing number. The nodes in Raft are hierarchically ranked in different states, with each node to be either a leader, a follower or a candidate. The leader is the principal entity of the protocol and it is elected per channel with the task to interact with the clients and then replicate its entries to its synchronised followers. Raft achieves the best possible synchronisation by sending systematic "heartbeats" to its followers and if the leader crashes, then at least one of its followers will detect this divergence, cast a vote to the network and attempt to take its place. Some nodes might compete to win the election process by seeking votes from other nodes. Therefore, these nodes are considered as "candidates".

Raft ensures exclusively that a single node will become a leader, even if some nodes miss a term or split ups occur in the leader election process. In the first case, the outdated node will revise the term's number and fall into the follower's state; and in the latter, the current term will be ended without any outcome of the election process.

Raft's performance is not fully tested, but its implementation in Fabric [65], as the recommended consensus protocol since the version 1.4.1, showed that it can provide thousands of transactions in real world scenarios, being two times faster than Kafka in terms of transaction's throughput with even less latency. In the context of Corda and Quorum, Raft reaches throughput less than 200 and 650 TPS with BCT as low as 1sec and 4.5sec, respectively.

### 6.6. Practical BFT (PBFT) protocol

The PBFT [41] is one of the most well-established BFT protocols and has recently attracted much attention with the emerge of the blockchain technology. PBFT works under the assumption that less than 33% of the network's nodes are behaving maliciously. Specifically, there are three phases required to establish consensus, namely the pre-prepare, prepare and commit. Each node is considered to be a *validating replica* (VR) that votes to elect a primary node. This primary node, called here *leader validating replica* (LVR), starts the three phase consensus, after receiving a request from a client, by multicasting a pre-prepare message. When the VRs receive this message, they enter to the pre-prepared phase and disseminate a prepare message to ensure that they are on the same sequence. When the LVR accepts $2f + 1$ prepare messages, enters to the prepared phase and multicasts a commit message. Therefore, there will be at least $2f + 1$ (as $N \geq 3f + 1$) correct nodes in the same state after the commit phase that produce the same result.

Despite PBFT's tremendous transaction throughput of 78000 TPS, measured in [28], the protocol has several drawbacks. The protocol lacks in terms of scalability, supporting only a few nodes and demanding several messages to be transferred to achieve consensus [11, 117].

### 6.7. Istanbul BFT (IBFT) protocol

IBFT [132] belongs to the family of *proof-of-authority* (PoA) consensus protocols and it is implemented in Hyperledger Besu [29] and Quorum [153]. The protocol leverages the potentials of smart contracts to ensure instantaneous finality and robustness in an eventually synchronous network. IBFT was initially designed as an alternative consensus protocol for the Ethereum blockchain that fits better with permissioned blockchains, in which the waste of the computational power to reach consensus is undesirable. The new version IBFT 2.0 of the protocol preserves all the original beneficial features of IBFT and addresses the security limitations, such as the security and the liveness properties. In addition, the number of the nodes that are required to reach consensus has changed and therefore, a byzantine node needs to reach the super-majority of the network, i.e. more than 2/3 of validators' votes, to insert a malicious transaction to the blockchain.

In terms of security, both versions of IBFT inherit their security properties from the three phase consensus. In contradiction to the similarities that IBFT and PBFT have, there are some tweaks that are needed to be made in the later to be used as a consensus protocol in blockchains. The fact that no particular client exists in IBFT to claim results by sending requests, considers all the VRs to be recognised as such and propose new blocks when their turn comes. IBFT's performance in Quorum is not peer-reviewed [21], but it is measured to be $600 - 650$ TPS with BCT of 4.5sec with very low scalability.

### 6.8. Clique / Aura protocols

Clique and Aura [11] are also PoA consensus protocols created to increase the platform's throughput, lower the confirmation time, but also to degrade the need for computational consumption in a permissioned setting. In the case of Clique, the time is separated into epochs, wherein the right to propose a block is given to the mining leader and other authorities only when a specific number of epochs has elapsed. In the case of Aura, each block is not committed at once, but instead it has to pass from the block acceptance round. If the participating authorities disagree on the leader's proposed block during the acceptance round, then the leader's position is under vote. Any leader can be considered malicious by any authority in the network only if: 1) it has not disseminated any block, 2) it has disseminated contrasting blocks to various authorities, or 3) it has disseminated multiple blocks. In any of the aforementioned cases, if the majority of the authorities agree, the malicious leader is removed from their list and all its proposed blocks are discarded. Clique is currently implemented in Besu [182], while Aura is implemented in Parity [148]. In terms of performance, the adoption of Clique in Besu is not yet measured, but Aura can achieve less than 50 TPS and BCT of 5sec. Despite their performance, both Clique and Aura are scalable protocols and can achieve consensus under the presence of byzantine nodes, with tolerance of $N \geq 2f + 1$.

### 6.9. BFT-SMaRT protocol

BFT-SMaRt [28] is a java-based library that encapsulates all the complexity of BFT algorithms in a simple API, which can be used to provide deterministic services. Apart from the modules that exist for trustworthy point-to-point communication, as a modular consensus protocol, it can provide reconfiguration modules that are completely separated from the process of agreement. These modules make the protocol vary from prior BFT networks, which are fixed-size networks that cannot be expanded or even reduced. In addition, some of its significant advantages are the features of multi-core awareness regarding the verification of signatures and the flexibility of the programming interface [176]. Compared to the existing SMR libraries, the protocol can reach better performance regarding not only the system's scalability, but also the transactions' throughput and latency, while providing the capability to withstand real-world failures that previous BFT implementations are not able to.

BFT-SMaRt is experimentally implemented in Fabric [9], Symbiont [180] and recently to SMaRtChain [27]. In Fabric's case, the protocol also implements WHEAT [176], a component that BFT-SMaRt's ordering service is relied upon to provide a powerful vote assignment scheme, low latency and fast replication among the VRs without compromising the network's security. The performance of BFT-SMaRt in Fabric is measured to be more than 10 thousand TPS with BCT of 0.5sec [176]. In the context of Symbiont, BFT-SMaRt is not fully tested, since it demands Java serialisation that is disabled by default due to security risks. Nevertheless, it can achieve the tremendous throughput of 80 thousand TPS with transaction latency even less than 1sec. The authors in [27] showed that, if the protocol is implemented in SMaRtChain, it can achieve high performance, even higher than the one achieved in Fabric - with a significant transactions' throughput of $12-13K$ TPS - and much lower latency, even less than 0.25sec.

### 6.10. VBFT protocol

As the core consensus protocol of the Ontology platform [145], VBFT [83] has been developed to address malicious behaviour via the combination of a PoS and a BFT protocol, while the randomness of the leader election process is achieved by means of *verifiable random functions* (VRFs). Therefore, the protocol obtains the best from all of these three worlds. In VBFT, the nodes are divided into two different types, namely, 1) the consensus nodes, which create blocks by weighting each node's stake and 2) the consensus candidates that only validate and update the consensus blocks, but they do not provide any further assistance in the consensus process. The VRF function is mainly used to select a difficult to be predicted type of work that each node will perform. The initiation of the protocol begins with each node to create and propose a new block. All the new blocks are then gathered, verified and voted by the verification nodes according to their highest priority. The results from the verification nodes are then validated from the confirmation nodes, which in their turn, establish a consensus result and when this new result is accepted, a new round begins. VBFT can withstand malicious behaviour, while implementing many features of scalability (such as, sidechains, sharding, and parallel execution). The protocol can deliver throughput more than 3K TPS and at peak performance 4 thousand TPS with BCT of $5-60$sec.

### 6.11. Delegated BFT (DBFT) protocol

DBFT [201] is currently implemented in Neo [139] and in Ontology [145] and as a combination of PBFT [41] and DPoS [111], includes the advantages of both worlds. To address malicious behaviour, the protocol follows the three-phase consensus, without inheriting PBFT's undesirable drawbacks. As more nodes join in the consensus process of the PBFT protocol, the faster its performance degrades, with time complexity of $\mathcal{O}(n^2)$ [139]. For this reason, DBFT adopts the characteristics of DPoS, in which the nodes with their voting power propose a class of delegates that will be authorised to establish consensus in the next round. Therefore, DBFT provides a fast voting scheme to elect the delegates and address any malicious behaviour. The beneficial features of DBFT regarding its scalability and its byzantine fault tolerance (as $N \geq 3f + 1$), come to contradiction with its performance metrics in the Neo platform, which are less than 1000 TPS and BCT of $15-20$sec.

### 6.12. Tendermint protocol

The fault tolerance properties of Tendermint [105] are established under a partially synchronous network for Hyperledger Burrow [67] and Ethermint [66], with high tolerance against double-spending attacks. The protocol relies on a leader election mechanism to maintain a constant consensus group, from which each validator's identities can be disclosed for public supervision.

Although Tendermint seems to be quite similar to the most of the BFT protocols, the voting power that the VRs use to elect a leader is defined in a different way and it is based on each VR's stake. In addition, a concept of "*polka*" or "*locks*" is added in the pre-commit phase to lock the blocks that collect more that the 2/3 of all the network's pre-votes. For various reasons, some honest blocks might fail to be included in a specific height; and therefore, to address this limitation, the protocol moves to a new round with a VR to propose a block for the new height. This block can be either a new block or a previous block, in which the proposer has been locked to from another round. The VRs can unlock from their blocks by collecting more than 2/3 of nil pre-votes. These type of votes, according on the phase that they occur, can define either that a VR has to be unlocked from a previous outdated block or that the protocol has to move to the next round. In any case, if the block is either locked (or unlocked) it has to be accompanied by a *proof-of-lock* (PoL), which is a collection of the (nil) pre-votes that it has received. Tendermint can provide medium performance metrics in Ethermint and Burrow, but the concept of locks has given the inspiration for the creation of another and high performance protocol; the Exonum [199].

In terms of performance, Tendermint can provide $200-800$ TPS and each block to be included in the blockchain in less than 1sec, while Burrow can achieve throughput more than 400 TPS and by default produces blocks every 2sec. Exonum on the other hand, shows significant results with transactions' throughput approximately at 5 thousand TPS and BCT of 0.5sec.

### 6.13. Proof-of-elapsed time (PoET) protocol

PoET [87] is designed to remove the computational power that PoW-based consensus protocols introduce. The protocol is implemented in Sawtooth [87] and solves the problem of Byzantine agreement by adopting a lottery-based method to achieve investment, verification and fairness in the leader election process. The peers wait for an indefinite time-period to be elapsed [8]; and the peer that finishes waiting first, is elected to be the leader and create a new block. This procedure is executed in a private memory area, called *trusted* *executed environment* (TEE), such as the Intel's SGX [48], which is mostly used in the Sawtooth platform. Essentially, the TEE provides a cohesion proof that is obtained from a trusted function via remote attestation techniques. This trusted function, which is often called "enclave", establishes trust on the consensus peers and allows the network to tolerate malicious behaviour. Each node calls an enclave inside the SGX to produce a random delay, wait for its timer to be finished, and then declare itself to be the leader. The protocol is BFT but without the integration of the SGX hardware or if the adversary manages to compromise a set of participating nodes beyond the $\Theta(\log \log n / \log n)$ threshold [44], the network can be manipulated. Although there is no any performance evaluation from the Hyperledger project, Ampel [8] showed that Sawtooth can achieve 2.3 TPS and latency less than 1sec.

### 6.14. Ripple protocol consensus algorithm (RPCA)

The Ripple consensus protocol is created to provide enhanced security and a stable real-time cryptocurrency-based network for transferring remittances without the obligation to develop smart contracts. In RPCA, each node maintains a *unique node list* (UNL), which is a set of trusted validators especially configured to participate in the consensus process. In Ripple's vast network, trust does not literally means that the set belonging to each node's UNL is actually trusted, but rather that this set is going to behave honestly and that the nodes included in each node's UNL are not intended to collude with each other or attempt to break the protocol's rules or manipulate the network with positive votes on malicious transactions. Each node listens to its trusted validators, but if agreement is not reached upon a set of transactions, the node's proposals are modified according to the proposals of their trusted validators. RPCA requires than 1/5 of nodes to be faulty in each node's UNL in order to achieve overall consensus. Therefore, only the transactions that exceed a specific threshold of 80% of positive votes are processed, while the rest are either deserted or included in a candidate set of new and not yet applied transactions on the ledger. Ripple can reach 1.5 thousand TPS and BCT in 4sec.

### 6.15. Yet another consensus (YAC) protocol

YAC [134] is based on voting for block hashes to solve the problem of inefficient message processing and strong leader election, that both occur in classical BFT consensus protocols. The novel design of YAC is created for the needs of Hyperledger Iroha to provide the simplest possible construction for the development of mobile applications [88]. Similarly to Fabric, the transactions' validation is not part of the consensus process, but it is rather determined by the platform's transaction flow. For each transaction to be included in a block, two validation operations are executed from the peers (a stateless and a stateful). The stateless validation ensures the validity of the transaction's format and signature, while the stateful occurs in a slower rate after the transactions' ordering. Iroha's operations are limited in ordering and consensus. When a transaction has passed from the stateless validation check, it is sent to the ordering nodes and if a number of transactions are accumulated or if a specific amount of time has elapsed, a batch that contains the proposed transactions is sent from the ordering service to the peers for the stateful validation. When the peers receive this proposal, they verify its contents by performing the stateful validation and then create a block containing only valid transactions. Afterwards, this block is voted through signing and if it receives more that 2/3 of of the total network's votes, the leader sends a commit message declaring that this block should be accepted to the peer's chains. In terms of performance, YAC can achieve high throughput of several thousand transactions per second, with each transaction to be confirmed in less than 3sec.

### 6.16. Stellar consensus protocol (SCP)

SCP [129] operates as a provably secure mechanism of federated byzantine agreement to address malicious behaviour across the Stellar network. The protocol is tailored towards financial purposes using quorums and quorum slices. Quorum is the set of nodes that attempt to establish agreement on each state's validity, once a specific threshold of nodes is met; and quorum slices are subsets of the quorum associated with and trusted by a given node, meaning that each node can rely on multiple slices to collect information. The SCP consists of two primary sub-algorithms, the nomination protocol and the ballot protocol. In each consensus slot, the nomination protocol is executed first, in which the new candidate values are produced and disseminated to all the quorum nodes that can vote only for a single one of them. With the termination of this process and the values to be unanimously selected for the current slot, the ballot protocol is initiated and includes a federated voting for either accepting or discharging the generated values that were created in the nomination protocol. In scenarios, where agreement cannot be reached (such as split votes), the nomination and the ballot protocol contain some high-level details to resolve these issues. For example, the execution of a higher valued algorithm that can serve the same purposes and be considered as the execution of a new ballot protocol.

In terms of performance, SCP caps out 1000 TPS with each transaction to be confirmed in $3-5$sec; and in terms of security, in a symmetric system with $N$ nodes and any set of $T > N/2$ nodes to constitute a quorum, a system designed to survive $f$ failures will be comprised of $N > 3f + 1$ nodes and a quorum size of $T = 2f + 1$.

## 7. Blockchain security

Blockchains are inherently secure by design, however, numerous recent security studies have proved that this new technology could be susceptible to various cyber-security attacks aiming at breaking its integrity and availability [64, 78, 112]. Just like any software program, blockchain possesses its own vulnerabilities that can be exploited by attackers to gain unauthorised access to the system and perform malicious operations [64, 78]. However, successfully exploited vulnerabilities in blockchain can have serious impact, as they can provide unauthorised access not only to the data stored at the compromised point but also to all data recorded on the distributed ledger, which could have wider negative consequences on the state of the blockchain. Some of these vulnerabilities are specific to particular blockchain platforms, while the others are general. In this section, we focus on the attack surface of the blockchain platforms whose consensus protocols were analysed in section 6, along with the security mechanisms that these platforms provide to address potential cyber-threats. Consensus-based attacks are not included in this evaluation as they are treated in section 6.

**Table 2**
Attacks on blockchains and indicative countermeasures

| | Blockchain attacks | Countermeasures |
|---|---|---|
| Identity | ■ Replay attacks<br>■ Key attacks<br><br>■ Impersonation attacks<br><br>■ Sybil attacks | ■ Elliptic curve encryption<br>■ Elliptic curve encryption<br>■ PKC key pairs' freshness<br>■ Elliptic curve encryption<br>■ Signcryption<br>■ Attribute-based signatures<br>■ New identity creation cost<br>■ Digital certificates |
| Service | ■ Refusal-to-sign<br>■ DDoS attacks | ■ Separation of transactions<br>■ ECDSA ring signatures<br>■ Block size limitation<br>■ Attribute-based signatures<br>■ Multi-receivers encryption<br>■ Distributed SDN architecture |
| Manipulation | ■ Tampering attacks<br><br>■ Overlay attacks<br>■ MiTM attacks<br><br>■ Eclipse attacks | ■ PKC signatures<br>■ Elliptic curve encryption<br>■ Timestamps for uniqueness<br>■ Elliptic curve encryption<br>■ Secure mutual authentication<br>■ Randomness in choosing peers<br>■ Secure channels |
| RP [1] | ■ Whitewashing attacks | ■ Low priority to newcomers |
| Crypto | ■ Quantum attacks<br>■ Brute force attacks | ■ QC / PQC techniques<br>■ QC / PQC techniques<br>■ Cryptographic keys' sizes |

[1] Reputation-based attacks

As illustrated in Table 2, cyber-attacks against blockchain platforms are grouped into the following categories, identity-based attacks, manipulation-based attacks, service-based attacks, reputation-based attacks, and cryptography-based attacks, which are analysed in the following subsections.

### 7.1. Identity-based attacks

In attacks under this category, the adversary aims to forge its identity, masquerade itself as a legitimate user, gain access to the targeted system and manipulate it [64]. Such type of attacks exploit weaknesses related to the security of the cryptographic keys. Any possible leakage of the cryptographic keys, can be used by adversaries to control the peer's identities and undermine the consensus process. The most interesting attacks in this category are the key, replay, impersonation, and the Sybil attacks. The replay attacks aim at spoofing the identities of two parties, intercept their data, and replay them to their destinations [64]. Such attacks occur due to the confusion that some nodes may experience after a hard fork and often within insecure cryptocurrency-based protocols. Various blockchain platforms, such as the Exonum [199], Bitcoin [165], Ethereum [194], Elastico [121] and Parity [148] are vulnerable to security attacks that might fork the blockchain [158]. Currently, Monero does not provide any security mechanism against the replay attack [114]. Corda [155], Quorum [153], Hedera [19], Ripple [167] and Ouroboros [94] proposed mitigation mechanisms, such as secure TLS mutual authentication and *elliptic curve cryptography* (ECC) to calculate the hash functions [106].

Key attacks exploit weak key schemes. Any inappropriate use or storage of the keys can allow unauthorised users to take control of the identities of the participating nodes. Especially, with the advance of the quantum computing, cryptographic keys can be easily cracked, leading to the unauthorised control of the victim's identity [64]. Similarly, in impersonation attacks, weak or leaked private keys can be exploited by adversaries to masquerade themselves as legitimate users and perform unauthorized actions in the system. For instance, they can be impersonated as "authorised block creators" and insert fake data to the blockchain. Unfortunately, the protection measures against these attacks in most blockchain platforms are not very robust [114]. For instance, Ethereum [194], Quorum [153], Parity [148] and Corda [155] are vulnerable to key compromisation attacks, since the account keys are generated using elliptic curve cryptography, which is proved to be weak against these type of attacks [106]. To deal with these attacks, Monero uses the One-Time Private Key (OTPK), which can be used only one time in conjunction with some well-established cryptography method [193]. IOTA replaced some uses of Curl hash function, which is vulnerable to signature forgery attacks, with a new hash function named Kerl [80].

In a Sybil attack, the adversary exploits leaked keys to create a large number of fake identities that can act as authenticated nodes and perform malicious interactions to intentionally increase or decrease the reputation scores of the targeted nodes [78]. BitShare, Lisk and EOSIO [85] cannot prevent this attack as an anonymous or pseudonymous party can participate in the validation and creation of blocks [78]; Tayebeh et. al. [156] found that Elastico is vulnerable to this attack. IOTA [89] is also prone to sybil attacks, since the malicious nodes can create random transactions to select a tip and thus, process invalid transactions. In Ripple, Stellar and Quorum the authorised participants must be authenticated to avoid a Sybil attack [158]. In blockchain platforms that implement a PoW-based consensus protocol, such as Litecoin [116], Monero and Ethereum [194] the malicious node, who wants to create a fake identity should divide its mining power to the corresponding number of sibyls and therefore, reduce its possibility to find the next block. However, Ethereum [194] is still vulnerable to Sybil attacks due to weak restrictions on the node generation process. Tezos [184] and Hedera [19] implement security mechanisms to avoid Sybil attacks, however, Hedera is considered to be prone due to the anonymous individuals that can participate in the network [19].

Exonum [199] has provided increased security against these attacks by using an anchoring service that writes the hash of the current Exonum state to the Bitcoin, in a certain time interval. The anchored data is authenticated by the majority of the validators using digital signatures. Therefore, even if the attacker compromises all the validators, it is impossible to change the transaction log unnoticeable. NEO [139] supports digital certificates, which can solve the potential problem of dishonest nodes in the public blockchain. In order to enhance keys' security, *hardware security modules* (HSMs) have been proposed to securely generate, protect, and store private and public keys [33].

### 7.2. Service-based attacks

This category involves attacks that aim to disrupt the normal operation of the blockchain network, making it unavailable for the legitimate users, or forcing it to operate in a different way from its specification [64]. There are two different types of attacks in this category and are: the *refusal-to-sign*, and the *denial-of-service* (DoS) or *distributed DoS* (DDoS) attacks [52, 64]. In a DDoS attack, the attacker typically leverages a number of hijacked devices to flood the network with an excessive number of requests that can disrupt the normal operation of the blockchain, degrade its efficiency, or make it unavailable for legitimate users. Most discussed platforms, including Bitcoin [165], Monero [141], Litecoin [116], Parity [148], Ethereum [194] and the Hyperledger frameworks, Fabric [9], Burrow [67], Iroha [88] and Sawtooth [87], are vulnerable to DDoS attacks, which become the most common attacks on the blockchain ecosystem [97, 78]. Particularly, blockchain platforms with a centralised structure are prone to this attack [51], e.g., Quorum (i.e., the structure of Quorum slices is highly centralised) and Stellar, which is highly depended on the structure of the quorum slices [95]. In the same context, a recent study found that over 30% of the users in EOSIO correspond to botnets, with more than 300 attacks that have already been detected [158].

In IOTA [15], the DDoS attack is prevented by requiring from the devices to issue a transaction and execute PoW, a requirement that needs only a few seconds, depending on each device specs; and ensures that a given device can not send multiple transactions in a short amount of time. Hedera [19] is also DDoS resilient, since it does not provide to any node or to any small number of nodes special rights or responsibilities in consensus process. Even if a DDoS attack floods a miner and temporarily disconnects it from the network, the community as a whole, will continue to operate normally. Liu et. al., [118] proposed a combination of ring-based signatures with the ECDSA in order to secure the transfer of coins between the user's addresses. Bitcoin and Bitcoin-NG include some protection against DoS attacks by limiting each block's size and the maximum number of the attribute signatures for the transaction input, but these platforms are still vulnerable to more sophisticated DoS attacks [97].

Refusal-to-sign attacks occur when a malicious node refuses to sign an unfavourable transaction. In this context, malicious nodes, especially in Bitcoin, may pollute the blockchain with refusal to sign attacks and therefore decrease the trust in the blockchain network. None of the aforementioned blockchain platforms can provide a concrete solution to this issue. A simple way to address this attack is proposed by TrustChain [146], in which the interactions that derive from such nodes are blocked and the transactions are split into smaller amounts.

### 7.3. Manipulation-based attacks

This category incorporates attacks that illegally attempt to intercept, manipulate or destroy sensitive data in transmit or at rest. Most common attacks are the overlay, the Man-in-the-Middle (MiTM), the eclipse and tampering attack [52]. Overlay attacks exploit blockchain vulnerabilities to maliciously enclose an encrypted amount to an original transaction by using the receiver's public key [64]. Protection against this attack can be guaranteed through the verification of the timestamps used to ensure uniqueness of the transactions. Therefore, the different inputs under the same trader can be identified and associated with the different transactions [64].

MiTM attacks exploit vulnerabilities like private key leakage to spoof identities of two parties and secretly intercept and even modify their communication, while they believe they communicate with each other in a secure channel [64]. Most blockchain frameworks include some form of endpoint authentication to prevent MiTM attacks. However, some of them are still vulnerable to this attack such as Ethereum and Bitcoin. For instance, Hedera, Quorum, Fabric, Corda, Sawtooth and Ouroboros platforms prevent this attack by using TLS mutual authentication among the nodes. All communication among Fabric nodes is configured to use TLS, however, it is still vulnerable to MiTM attacks and SSL Stripping, which is a type of an MiTM attack, due to the centralisation of the MSP admins and the OSNs [51]. In IOTA [80], messages are transferred over a TLS-encrypted TCP/IP connection with mutual certificate controls to prevent these attacks. On the other hand, Exonum [199] includes a lightweight client that provides additional security against MiTM attacks and maliciously acting nodes. Tezos [184] also supports secure connections via TLS v1.3 and v1.2 with forward secrecy based on elliptic curve Diffie-Helman key exchange. However, Bubberman et al. [39] found that it is possible to conduct a TLS MiTM attack on the Ripple XRP ledger to gain access to message content.

Eclipse attacks attempt to isolate the victim node by monopolising all of the its incoming and outgoing connections. This allows the attacker to corrupt the victim's view of the blockchain, force it to waste computing power or leverage the victim's computing power to conduct its malicious acts. Eclipse attacks can trigger other attacks like Selfish mining, 51% attacks, etc. [114, 52]. Research works [81, 192, 124, 78] proved that Bitcoin (*see* Figure 3), Litecoin, Bitcoin-NG, Monero and Ethereum are vulnerable to this attack. However, some security defences have been taken to deal with the eclipse attack, but these defences are rather inadequate. For instance, Bitcoin [165] enhanced the randomness by choosing IP address from the tried table. Work by Zhang et. al. [202] have also demonstrated the feasibility of launching Eclipse based stake-bleeding attacks against blockchain systems that are based on PoS like Snow-White and Ouroboros. This attack is a combination of the Eclipse and the stake-bleeding attack. The combination of a virtual voting with a gossip-to-gossip protocol in Hedera is not adequate to provide resistance, but by using secure communication channels its impact can be reduced. In addition, implementing defences against Sybil attacks can bound the portion of overlay nodes controlled by the attacker [171].

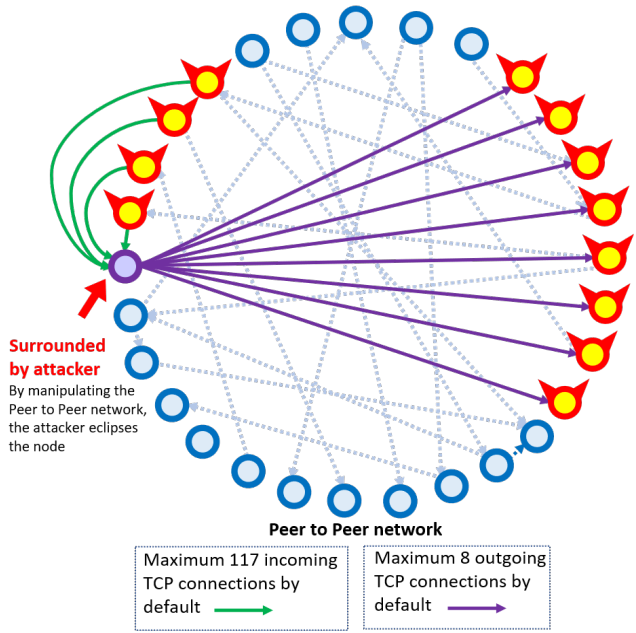

**Figure 3:** Eclipsing a Bitcoin node: The nodes have a limited number of outgoing connections, therefore, the victim is forced to establish connections only with the malicious nodes that were created by the attacker [52].

Table 3
Weights assigned to the attacks considered.

| $w_{ik}$ | Attack ($\text{Att}_{ik}$) | $w_{ik}$ | Attack ($\text{Att}_{ik}$) |
| --- | --- | --- | --- |
| 0.30 | Replay attacks | 0.20 | Key attacks |
| 0.25 | Impersonation attacks | 0.25 | Sybil attacks |
| 0.50 | Refusal-to-sign attacks | 0.50 | DDoS attacks |
| 0.20 | Tampering attacks | 0.20 | Overlay attacks |
| 0.40 | MiTM attacks | 0.20 | Eclipse attacks |
| 1.00 | Whitewashing attacks | | |

Tampering attack is one of the main security risks to the blockchain platforms. In this attack scenario, the attackers attempt to modify the signed transactions being disseminated in the network, the addresses and other data, and then propagate the tampered transactions to the peer-to-peer network for validation. Bitcoin [165] is prone to this kind of attacks. Most discussed platforms try to follow NIST's guidelines for using cryptographic standards to avoid weak and non-secure schemes in order to prevent such attacks. In this context, Paillier encryption scheme has been proposed to preserve the Bitcoin transactions privacy and hide the plaintext amounts in the transactions [190]. Tampering attacks could also delay the propagation of transactions and blocks to specific nodes [47]. PoW cryptocurrency-based ledgers like Bitcoin, Litecoin and Monero are more vulnerable to this attack [78]. Arthur et al [68] found that attackers can provoke significant delays in the broadcast of Bitcoin packets by introducing congestion in the network or making a victim node busy by sending requests to all its ports.

### 7.4. Reputation-based attacks

In this category, attackers attempt to manipulate the reputation of specific nodes in the blockchain network by intentionally decreasing or increasing their reputation. In this context, a whitewashing attack is a big concern for the blockchain community. In this attack, nodes with poor reputation, exploit some system's vulnerabilities to rejoin the blockchain with new identities and update their reputation. Currently, there is no concrete solution against this attack [52], only TrustChain [146] provides lower priorities and capabilities to nodes with new identities.

### 7.5. Cryptanalytic attacks

The robustness of blockchain platforms highly depends on the public-key cryptography and the hash functions, such as the RSA encryption scheme and the elliptic curve cryptography. However, using mathematical encryption methods could no longer guarantee the security of these platforms, especially with the fast progress of quantum computing. This technology will allow the implementation of algorithms that can efficiently solve classes of mathematical problems that cannot be practically solved today. In fact, a quantum computer is able to break most of the current encryption algorithms. With these features, brute-force attacks will only take a few minutes to crack the encryption and easily retrieve the encryption keys. As illustrated in Table 6, some blockchain platforms like Bitcoin-PQ, IOTA and Corda have implemented quantum and post-quantum cryptographic techniques to tackle the quantum threat. Some other platforms, such as NEO, Algorand and Fabric are currently transitioning to a quantum resistant infrastructure. However, in a very recent implementation [82], a hybrid quantum-safe Fabric network has been presented, in which hybrid digital signatures are incorporated and provide security against both classical and quantum attacks. Quantum cryptography uses the principles of quantum mechanics to encrypt data and make it impenetrable, while, post-quantum cryptography refers to the security against an attack by a quantum computer.

### 7.6. Synopsis

Next, we provide a summary of the analysis carried out in the previous subsections and derive an overall security score (*see* Table 6) that expresses the ability of the blockchain platforms to mitigate attacks, in each of the aforementioned categories. In particular, the score $\text{sec}_j$ achieved by platform $j$ is given by the following expression

$$\text{sec}_j = 0.35 R_{j1} + 0.20 R_{j2} + 0.35 R_{j3} + 0.10 R_{j4} \quad (1)$$

where $R_{ji}$ is the resistance score of the $j$th platform against the $i$th attack category; that is, the index $i = 1, \ldots, 4$ is associated with *identity*, *service*, *manipulation* and *reputation* based attacks. Due to the different means of performing cryptographic attacks, these are separately considered in Table 6 and the overall comparative evaluation of section 10. The weights given in (1) were based on the estimated impact of each attack category; for example, the impact of manipulation-based attacks (e.g. tampering attacks) is high, as it affects the integrity of the data eventually stored in the shared ledger. The value of $R_{ji}$ is obtained via

$$R_{ji} = \sum_k w_{ik} S_j(\text{Att}_{ik}), \quad i = 1, \ldots, 4 \quad (2)$$

where $w_{ik}$ is the weight of attack $\text{Att}_{ik}$ within the $i$th attack category and $S_j(\cdot)$ is a discrete-valued function quantifying the security score achieved by platform $j$ against a specific attack. The weights assigned to each attack are summarised in Table 3 (again, excluding cryptanalytic attacks), while the possible scores given by function $S$ are defined as

$$S_j(\text{Att}) = \begin{cases} 0.1, & \text{if platform is vulnerable to Att,} \\ 0.5, & \text{if platform has defences against} \\ & \text{Att but are rather inadequate,} \\ 1.0, & \text{if platform is robust against Att.} \end{cases} \quad (3)$$

The obtained security scores $\text{sec}_j$ for all platforms (i.e. all $j$) are presented in Table 6. However, to ease presentation, two typical examples are presented next, to clarify how each platform's overall security scores is calculated and a three-level scale *low*, *medium*, and *high* has been used; the level for each platform is computed as follows.

$$L(\text{sec}) = \begin{cases} \text{Low}, & \text{if sec} < 0.3, \\ \text{Medium}, & \text{if } 0.3 \leq \text{sec} \leq 0.7, \\ \text{High}, & \text{if } 0.7 < \text{sec.} \end{cases} \quad (4)$$

Therefore, *high* level indicates that the blockchain platform offers a complete security solution to address most of the attacks discussed; *medium* level is assigned to the platforms addressing a sufficient subset of the attacks, while *low* indicates that the blockchain under consideration is vulnerable to most of the attacks discussed here.

**Example 1.** Algorand, is not found to be robust against any of the aforementioned attacks, but vulnerable to sybil, refusal to sign and whitewashing attacks. The platform also provides defences against replay, key, impersonation, DDoS, tampering, overlay, eclipse and MiTM attacks, but these defence mechanisms are rather inadequate. Hence, we need to calculate (2) for $j = 1$, which corresponds to Algorand; the scores obtained for each attack category ($i = 1, \ldots, 4$) are

$$R_{11} = w_{11} 0.5 + w_{12} 0.5 + w_{13} 0.5 + w_{14} 0.1 = 0.4$$
$$R_{12} = w_{21} 0.1 + w_{22} 0.5 = 0.3$$
$$R_{13} = w_{31} 0.5 + w_{32} 0.5 + w_{33} 0.5 + w_{34} 0.5 = 0.5$$
$$R_{14} = w_{41} 0.1 = 0.1$$

Hence, evaluating (1) for Algorand gives the overall resilience score of

$$\begin{aligned} \text{sec}_1 &= 0.35 R_{11} + 0.20 R_{12} + 0.35 R_{13} + 0.10 R_{14} \\ &= 0.39 \end{aligned}$$

which corresponds to a medium level of security from (4) ∎

**Example 2.** Fabric, on the other hand, is found to be robust against tampering and overlay attacks, as well as, vulnerable to refusal to sign, DDoS and whitewashing attacks. The platform also provides defences against replay attacks, key attacks, impersonation, sybil, eclipse and MiTM attacks, but these defence mechanisms are rather inadequate. Hence, we need to calculate (2) for $j = 2$, which corresponds to Fabric; the scores obtained for each attack category ($i = 1, \ldots, 4$) are

$$R_{21} = w_{21} 0.5 + w_{22} 0.5 + w_{23} 0.5 + w_{24} 0.5 = 0.5$$
$$R_{22} = w_{31} 0.1 + w_{32} 0.1 = 0.1$$
$$R_{23} = w_{21} 1 + w_{22} 1 + w_{23} 0.5 + w_{24} 0.5 = 0.7$$
$$R_{24} = w_{41} 0.1 = 0.1$$

Hence, evaluating (1) for Fabric gives the overall resilience score of

$$\begin{aligned} \text{sec}_2 &= 0.35 R_{21} + 0.20 R_{22} + 0.35 R_{23} + 0.10 R_{24} \\ &= 0.45 \end{aligned}$$

which corresponds to a medium level of security from (4) ∎

The analysis shows that blockchain 1.0 and 2.0 platforms (e.g. Bitcoin [165], Ethereum [194], Litecoin [116], etc.) have relatively *low* security levels as they are subject to a number of security attacks such as DDoS, sybil, replay, eclipse, MiTM, and tampering attacks. Study in [60] affirmed that several currency exchanges have been shut down as a result of DDoS attacks. In addition, the Sawtooth [87] platform lacks adequate security mechanisms against the considered attacks. Most of the surveyed blockchain platforms achieved *medium* security levels, since they adopt some level of protection mechanisms against the aforementioned attacks. However, none of these mechanisms is adequate enough to address them. For instance, the blockchain platforms that implement the DPoS consensus protocol, such as BitShares [32], Lisk [115], Tezos [184] and Tron [184] are vulnerable to centralisation, while the risk of DDoS, sybil and eclipse attacks is increased due to the strictly limited number of witnesses. Amongst the discussed platforms, Hedera [19] achieves *high* security level, as it provides protection against most of the considered attacks [79]; however, its security properties need further investigation.

## 8. Smart contracts' security

Security is a critical requirement in the context of a smart home. Within such a resource restrained environment it is excessively difficult to enforce security by increasing the key sizes with multiple cryptographic operations. The IoT devices are restricted in the form of storage capabilities and computational power. Therefore, the use of digital public key certificates can be an expensive operation for the smart home's IoT devices. The systems based on *public key infrastructures* (PKI) might demand verification via the requests sent to the cloud servers which will create the network's traffic. In this case, the privilege definition and the access control on the centralised servers will be vulnerable.

Blockchain technology can provide the functionality of executable logic via smart contracts. This functionality can be used to handle the interactions of a smart home ecosystem as transactions, by issuing commands to the IoT devices and ensure confidentiality. The smart home IoT devices can trigger the smart contract functions when some specific conditions are satisfied and call them with addresses or prompt them as application reaction to listening events. Therefore, the smart contracts can execute security operations, like authentication and access control, as well as, to enhance the overall security of a blockchain-based smart home ecosystem. Within this ecosystem, the service provider and the IoT data owner can interact as transacting entities without a trusted third party. With the use of smart contracts, the granularity of the IoT data being shared is in the control of the data owner, who can determine exactly the amount, the type and the time-span of the data to be shared. Therefore, smart contracts can be used to grant or revoke access privileges and safeguard the home-owner's privacy.

### 8.1. Smart contracts' vulnerabilities

Upon its invocation, the smart contract code is stored on the blockchain, while the smart contracts' operations can be called at any time by any participating node. The smart contracts are often called as autonomous agents, due to the fact that they are being given their own accounts on the network and their own network addresses. Therefore, their implementation can hold ownership of tokenised assets or custody, while the engaging parties are complied with the agreed-upon conditions. Their invocation often incurs an execution fee, since each operation is considered as a transaction logged into the blockchain platform. The smart contracts can be used to execute a variety of operations within the blockchain network, such as to: *a)* enable multi-signature transactions, where each transaction is only executed when the majority of the participating entities sign it; *b)* enable automated transactions generated by unique events; for instance, transactions that are automatically disseminated over specific time periods or transactions that are disseminated in response to other transactions; *c)* provide services to other smart contracts; for instance, in occasions where a smart contract invokes operations from another smart contract, and *d)* allow storage space for application-specific information, such as Boolean states, lists, membership records, etc.

**Table 4**
A taxonomy of vulnerabilities in smart contracts [16].

| | Vulnerability | Cause |
|---|---|---|
| source code | Call to the unknown | The called function doesn't exist |
| | Out-of-gas send | Fallback of the callee is executed |
| | Exception disorder | Exception handling irregularity |
| | Type casts | Contract execution type check error |
| | Re-entrance flaw | Function re-entered before exit |
| | Field disclosure | Private value published by peer |
| EVM[1] | Immutable bug | Contract change after deployment |
| | Ether lost | Send ether to orphan address |
| | Stack overflow | Number of values in stack > 1024 |
| blockchain | Unpredictable state | State modified prior invocation |
| | Randomness bug | Seed biased by malicious peer |
| | Timestamp dependence | Timestamp modification by a malicious peer |

[1] Ethereum virtual machine bytecode

Though it is hard to modify records stored on a distributed ledger, it is possible to compromise the software systems implementing the technology; the hack of Mt. Gox, resulting in $450 million losses, is a notable example. Another incident is related to the *decentralised autonomous organisation* (DAO), holding a large percentage of Ether; it suffered ∼$60 million in losses when a smart contract vulnerability was exploited leading to an infinite recursive calling situation that blocked the invocation of the function updating a user's balance. A summary of vulnerabilities in smart contracts is provided in Table 4. Most of these vulnerabilities apply to Solidity, which is an object-oriented, high-level language programming language supported by Ethereum [16]; this is due to the misalignment between the language semantics and programmers' intuition, and the fact that Solidity does not have a structure to deal with domain-specific attitudes in public blockchains leaving them open to reordering or delaying. Yet another cause for the insecurity of smart contracts is the lack of a single source of documentation for known vulnerabilities, being comprehensive, self-contained and regularly updated, so as to avoid past mistakes.

### 8.2. Smart contracts' verification

As a result of the severe consequence that can occur from the insecurity of the smart-contracts, researchers have adopted several techniques and tools that detect these vulnerabilities. Theorem proving, symbolic execution, model checking and abstract interpretation can be used to improve the security aspect of smart contracts. Therefore, in this section, we investigate methods and techniques that have been used for their verification and their improvement. The characteristics of these frameworks and tools are summarised in Table 5. The majority of them choose static analysis and support

EVM bytecodes analysis.

### 8.2.1. Theorem proving

*Theorem proving* is the most ordinary method to formalise smart contracts. The system is modelled mathematically, while the aimed properties to be proven are stated, with the verification to be executed on a computer by a theorem prover software. The theorem prover applies well-known axioms and simple inference rules for the verification process with each new theorem or lemma that is required for the proof can be derived from them. Theorem proving can be considered as an adaptable verification method, due to its suitability on all the systems that can be expressed in a mathematical way.

Theorem provers can be automated, interactive or hybrid [77]. The automated theorem provers perform the proving automatically, while the interactive might need some human input. Whether it is going to be used automated theorem proving or interactive, it is depended on the system's complexity, but for both, manual set up is required for the specifications and the system model. The hybrid theorem proving allows the partitioning of the system's model based on the complexity level. Then most complex parts are verified with interactive theorem proving, while the least complex parts are verified with automated. For some simple systems, statement logic or propositional logic can be adequate to verify the smart contract.

The most expressive logic for theorem proving is *higher-order logic* (HOL), which enables the use of quantifiers over sets and functions. Several frameworks also exist in the context of the smart contracts. The authors in [7], used an existing EVM-formal model and the logical framework Isabelle /HOL to propose a logic for EVM bytecode. The authors in [175], used formal verification of smart contracts with the K-Framework for expressing the EVM bytecode using logic. They created a simple smart contract on the Ethereum platform, wrote the high-level and the EVM-level specifications, and then verified this smart contract. Ether-Trust [75] is the first static analyser, which is developed by Grishchenko et. al., to detect the re-entrancy vulnerability. FSPVM-E [198], which stands for *formal symbolic process virtual machine*, verifies the security and reliability of Ethereum-based services at the source code level of smart contracts. This verification system is formulated in Coq and is comprised of four primary components: 1) the formal memory model (denoted as GERM); 2) the formal specification language for Solidity (denoted as Lolisa); 3) the execution and proof engine (denoted as FEther) and 4) the assistant tools and libraries that enhance the automation's degree and the validation efficiency of the system. $F^*$ [179] is a high-level general-purpose programming language, which is built for formal verification. The authors in [30], made an effort to formally verify Ethereum smart contracts, by translating the contract code to $F^*$.

**Table 5**
A taxonomy of verification methods for smart contracts

| | Method | Framework | Level |
|---|---|---|---|
| smart contract verification | theorem proving | ■ Isabelle/HOL<br>■ K - Framework<br>■ F* - Framework<br>□ FSPVM-E<br>■ EtherTrust | Bytecode<br>Bytecode<br>Bytecode & Solidity<br>Solidity<br>Bytecode |
| | model checking | ■ Zeus<br>■ VERISOL<br>□ VeriSolid<br>□ ContractLARVA | Solidity<br>Bytecode & Solidity<br>Solidity[1]<br>Solidity |
| | symbolic execution | ■ Oyente<br>■ teETHER<br>□ MAIAN<br>■ Gasper<br>■ sCompile<br>■ Osiris<br>■ Mythril<br>■ Slither | Bytecode<br>Bytecode<br>Bytecode<br>Bytecode<br>Bytecode<br>Bytecode<br>Bytecode<br>Bytecode |
| | abstract interpretation | ■ MadMax<br>■ Securify<br>■ Vandal<br>■ EthIR<br>■ EtherTrust | Bytecode<br>Bytecode<br>Bytecode<br>Bytecode<br>Bytecode |

■ Static analysis is performed in a non-runtime environment and without executing the contract, but by examining the source code for signs of security flaws.
□ Dynamic Analysis is performed while the contract is executed by examining it and trying to manipulate it in order to discover security flaws.
[1] as transition system

### 8.2.2. Model checking

*Model checking* is an automated technique for formal verification, applicable for systems that can be expressed by a finite-state model. The verification is performed using a model checking software, such as NuSMV. The user provides the model checker with a finite-state model of the system and a formal specification that it should possess. Then the model checks automatically if every state of the model satisfies the given specification. For the model to be formally verified for a specific property each state of the model must satisfy the specification [135]. If the specification is not satisfied, then a counter-example is given by the model checker, which shows a run of the system that violates the specification. This can help the user to identify mistakes and to correct bugs.

Model checking has been mostly used on Solidity, but it can verify contracts written in other languages, as well. ZEUS [53] is an automatic framework for formal verification by using abstract interpretation and symbolic model checking to soundly reason about program behavior. The smart contracts, which are written in high-level languages, are given as input to the ZEUS framework that uses the user's assistance to help generate the correctness or/and the fairness principles in a XACML-styled template [173]. *Verifier* *for Solidity* (VERISOL) is a formal verifier for smart contracts (written in Solidity), while it is not tied to Workbench. VERISOL encodes the semantics of the smart contracts into a low-level intermediate verification language Boogie [22] and leverages the Boogie verification pipeline not only for verification, but also for counter-example generation. VeriSolid [128] is built on top of FSolidM [127] and WebGME [125] and provides an approach for correct-by-design development of Ethereum-based smart contracts.

Model cheking is also well adopt in behavior-based verification. In [1], the authors based on a formal model checking language, proposed a novel way to model smart contracts, which was a way to model the users' behavior. In [56], Ellul at. al. proposed a runtime verification method to ensure that all execution paths followed at runtime, can satisfy the required specification, which are enclosed in the proposed method, called ContractLarva.

### 8.2.3. Symbolic execution

*Symbolic execution* is a software testing technique that helps to test data generation and proofs regarding the quality of a given program [20]. This technique is a delicate solution, in which possible execution paths are systematically explored simultaneously, without necessarily requiring concrete inputs. Instead of taking on fully specified input values, this technique represents them as symbols, in an abstract way, resorting to constraint solvers to construct actual instances that could cause property violations. Symbolic execution leads to major practical breakthroughs in a number of prominent software reliability applications and it is considered as the most prominent technique for vulnerability detection in EVM bytecode.

Several tools have been proposed that use symbolic execution. OYENTE [120] aims at timestamp dependence, mishandled exceptions, transaction-ordering dependence, as well as, re-entrancy bugs. The tool was used to check the first $1,459,999$ blocks of Ethereum and flagged $8,833$ contracts as vulnerable, out of the 19.366, including the DAO bug. teETHER [104] automatically generates transactions that can exploit the given contracts. MAIAN [140] is a dynamic tool that analyses the invocations of smart contracts to identify three types of trace vulnerabilities: prodigality, greediness and suicidality. Chen at. al. [45] developed Gasper [45], a static analysis tool that detects gas-costly patterns instead of vulnerabilities. sCompile [42] reports program paths containing monetary transactions, called critical paths; it discovered 224 unknown vulnerabilities with a false positive rate of 15.4%. Osiris [185], Mythril [137] and Slither [61] are some additional examples that use symbolic execution.

### 8.2.4. Abstract interpretation

*Abstract Interpretation* [49] a basic static analysis technique that soundly estimates program semantics. Various analysers have used this technique for EVM bytecode, since it can examine any possible execution. MadMax [73] identifies three gas-related vulnerabilities: non-isolated external invocations, unbounded mass operations and integer overflows. For an input property and a given smart contract SECURIFY [187] checks the smart contract behaviors with respect to the property. The analyser works in two phases: first precise semantic information from the code is extracted and then the compliance and violation patterns are checked to prove if a specific property holds. Vandal [35] leverages the Soufflé language [91] to generate results for custom vulnerability queries. EthIR [3] is a static analyser for Ethereum bytecode that relies upon an extension of the Oyente tool. ZEUS [53] and EtherTrust [75], mentioned earlier, are frameworks that also use abstract interpretation.

## 9. Blockchain privacy

In principle, a public blockchain platform raises more privacy concerns than a private. Similarly, in the latter case, it is easier to cope with privacy issues in a permissioned setting than in a permissionless one [149]. However, the form of the platform is actually contingent on the specific application; in other words, each application determines, by its nature, which type is more appropriate. For example, in the discussed case of integrating the blockchain technology to a smart home ecosystem towards facilitating and enhancing security services, the home-owner's personal data (including user's device identifier data) should not be public and, moreover, any access on them should satisfy the data protection requirements (i.e. on the basis of the user's explicit informed consent or some other legal basis). Clearly, in this case, a private permissioned blockchain network takes precedence over the other approaches. On the other side, other privacy requirements occur in other cases (e.g. in cryptocurrencies offered to the public, in which anyone can contribute in adding transactions to the ledger, a private blockchain network is not an option - and, hence, this in turn poses other data protection requirements). Blockchain technology can also be used to enhance the transparency of the user's communication, via providing an audit trail of the underlying data processes; again, in such scenarios, other data protection requirements may be present.

In any case though, several privacy enhancing mechanisms exist, to alleviate privacy issues arising by the use of the blockchain technology as the means to process personal data. As already stated in section 4.3, the necessity of whether each of these technologies should be implemented or not can be assessed only on a case-by-case basis, since each of them focuses on a specific privacy threat which may or may not be present, depending on the specific application and implementation decisions. We next present these techniques, with a discussion to the relevant blockchain platforms in which are being met.

### 9.1. Pseudonymisation

Privacy mechanisms on pseudonymisation are available in all of the blockchain platforms shown in Table 6 and these mechanisms are used to protect individuals identities and allow them to retain some utility on their data. According to the GDPR, pseudonymisation is the processing of personal data, in such a manner that the personal data can no longer be attributed to a specific individual without the use of additional information, provided that such additional information is kept separately and is subject to technical and organisational measures to ensure that the personal data are not attributed to an identified or identifiable natural person. In typical scenarios, pseudonymisation is achieved by replacing users identifiers by pseudonyms - i.e. specific type identifiers that do not allow, by themselves, re-identification of the individuals.

Blockchain platforms that are being used for cryptocurrencies (e.g. Bitcoin, Ethereum) employ pseudonymisation techniques by default, since each user having a wallet is uniquely associated with a random-looking address (which is generated through cryptographic mechanisms and plays the role of a pseudonym). Therefore, it can be seen that in public financial transactions, amounts of cryptocurrency being sent by one address to another, whereas there is no direct way to map the address back to an identified individual (typically, these addresses are derived by cryptographic public keys). In a similar way, any blockchain platform, for any application, may store pseudonymous data on the ledger, if this is necessary with respect to personal data protection requirements within a specific context; such a pseudonymisation may take place offline, so as to assign pseudonyms which in turn are being used in any blockchain transaction (e.g. pseudonyms may be assigned to user's devices identifiers in a smart home ecosystem). This offline process allows for providing the additional information required to reverse the pseudonymisation, if such a pseudonymisation reversal is needed.

It should be pointed out though that pseudonymous data are still personal (and not anonymous) data. Moreover, depending on the pseudonymisation technique and other information being available, a third party may manage to identify an individual by her/his pseudonymous data. For example, in the case of the classical Bitcoin, it is known that a former federal agent managed to trace 3760 bitcoin transactions over 12 months [113], finding out the person behind a specific Bitcoin address (i.e. pseudonym). As also stated in [113], the possibility - under specific preconditions - of tracing back the Bitcoin transactions to an individual exists due to the facts that: i) it is a public blockchain, ii) the public address of each user is always the same (public key-based address).

### 9.2. One-time address

To alleviate some issues arising from the usage of one pseudonym per user, a user may be associated by different pseudonyms (or addresses) to prevent linking transactions corresponding to a single user. To this end, using one-time addresses will quickly result in a large number of addresses per participant. Privacy mechanisms relying on one-time address are available in Bitcoin, Ethereum, Monero, as well as, in Ripple.

Such a characteristic example in blockchain technology is the use of the so-called *stealth addresses*; this notion has been first introduced for financial blockchain platforms, aiming to protect payee's privacy (as in the case of the Cryptonote technology [177]). More precisely, by appropriately using cryptographic mechanisms, such as the Elliptic Curve Diffie-Hellman algorithm, the payer creates a one-time address for every transaction with a specific payee, in order to enhance unlinkability for the latter. This is the case of Bitcoin. Apparently, techniques of this form could be also employed in other types of blockchain platforms, if such an unlinkability is needed; Ethereum and Monero (the latter being based on the Cryptonote technology) support one-time addresses.

In the case of Ripple, the notion of the so-called blinded tags has been introduced; each such tag consists of a sequence of numbers corresponding to the recipient of a transaction that is meaningful only for the intended recipient and is random-looking for everyone else.

### 9.3. Mixing techniques

Privacy mechanisms relying on Mixing techniques are available in IOTA, Ethereum and Bitcoin; this technique allows multiple participants to shuffle multiple transactions without revealing the exact relationships between the transactions - i.e. they cannot be linked. Again, such a technique was first developed to address privacy issues in cryptocurrencies. The idea of mixing is quite old [43], having used in several contexts. In blockchain, mixing techniques are used to conceal the history of a particular token (see, e.g. [163]), since each resulting (mixed) transaction that is stored on the blockchain corresponds to multiple senders and receivers - and, thus, rendering difficult for an adversary to single out one transaction. Mixing services (also known as tumblers) could be provided either through a centralised mixing service provider or on a peer-to-peer basis. For the Bitcoin, the Mixcoin protocol has been introduced in [34], being a centralised mixing service, whilst several other mixing services being compatible with the Bitcoin have been also proposed (a recent survey can be found in [151]).

It should be stressed though that the information (transactions) recorded in the ledger is in a readable/interpretable form, when mixing is being used and, thus, some linkability may occur in specific cases.

In the IOTA platform, there is is a coin-mixing service being called TangleMixer, with the aim to hide the sources of the transactions. The Ethereum also supports mixing services, like the MixEth [168] and Möbius [130].

### 9.4. Ring signatures

Ring signature is a special type of a cryptographic digital signature; this technique is only available in Monero and provides the following property. Given a group of users (members), a third party can verify that a signature has been put by a member of this group, but she/he cannot discriminate exactly which member is the actual signer [161]. Ring signatures are being considered as a nice privacy enhancing technology in blockchains, in terms of hiding the initiator of a blockchain transaction.

Similarly though to the case of mixing, the transaction information that is stored in the ledger is still in a readable form and thus, in specific cases, this may allow singling out of a signer.

An improved version of the ring signatures yields the so-called ring confidential transaction, which obfuscates the actual amount of the transaction that is being sent from the sender to the receiver. Therefore, only the participants in the transaction are able to see its amount, whilst, at the same time, the network is able to confirm the validity of this transaction. Such a technique is being called as Multi-layered Linkable Spontaneous Anonymous Group signature (ML-SAG) and is currently being used in the Monero platform [141] - which is a cryptocurrency based on the Cryptonote technology. However, it seems that MLSAG has not been used yet in the context of IoT applications.

A somehow generalisation of ring signatures is the multi-signatures, which allow multiple signers to jointly authenticate a message using a single compact signature [23]. Although the technology is not new, it has become widespread during the last years in the world of cryptocurrencies. Multi-signature technology was first implemented in Bitcoin since 2012. For example, Ethereum, IOTA, Hyperledger Iroha, Stellar, Tron, Ripple and Algorand blockchains support multi-signatures. However, it should be pointed out that multi-signatures mainly focus on enhancing security rather than privacy, since that the private keys needed to sign/authorise can be spread across multiple machines, eliminating any one of those machines as a single point of failure.

### 9.5. Confidential transactions

Several tools exist to achieve this technique in a blockchain platform; i.e enabling participants to hide their transaction data from unauthorised third parties (either other legitimate users of blockchains or not), while still the transactions can be executed. Note that this is something different from the aforementioned ring signatures, in which the contents of the transactions are being shown to all. Privacy mechanisms relying on confidential (or private) transactions are available in Quorum, Hyperledger Besu and Sawtooth, Symbiont, Ethereum (Parity), Ouroboros Crypsinous and Bitshares.

In Quorum, an optional feature of the so-called private transactions exists, allowing a user designate parties to privately transact with and, subsequently, the public ledger stores the hash value of the transaction whereas the original data are stored into private blockchains of the authorised participants. Therefore, all participants share a common public state which is created through public transactions, whereas they also have a local unique private state. If the transaction is private, the participant can only execute the transaction if it has the ability to access and decrypt the payload. Participants who are not involved in the transaction do not have the private payload at all. Hence, a private transactions manager serves as a off-chain privacy mechanism (related with the case of off-chain transactions, described next). Quorum communicates with the private transaction manager using HTTPS and keeps a reference to private transactions with relevant state trees on the blockchain.

In the Hyperledger Besu, a Private Transaction Manager is being used, to ensure that only involved parties can access the contents of transactions. In Hyperledger Sawtooth, there exists the concept of the so-called private-transaction-families, enforcing a policy of access control to the ledger. As stated in [87], this solution supports encryption of information in transactions and blocks while allowing the ledger to validate the information in those transactions in all the nodes and allow them to reach a consensus on the current state of the ledger.

In the Symbiont ledger, all the data is on the ledger but encrypted such as - despite the fact that all the data is stored on all nodes - a node can only decrypt what it is allowed to see, and likewise, it can just execute transactions that it's able to decode [180].

Parity, which is an Ethereum client supporting standard Ethereum APIs and protocols Technologies, implements a private transaction and smart contract capability in which a private contract's code and state are stored encrypted inside a public contract. The private transaction capabilities sit on top of Parity's existing Parity Ethereum Client [162].

In Bitshare, there is the notion of blinded transfers, in order to hide the balance amount and the identity of the party that is currently in control of the balance. But the transactional history of that balance may reveal details of the balance that can be deduced by inference.

Another way to hide transaction information is through cryptographic commitment schemes, such as the Pedersen commitment [150]. By these means, a sender creates a commitment to a transaction information and share the commitment instead of the information (see, e.g., [26]). The sender cannot lie later on on a fake value of the initial information (which remains though hidden). In financial transactions (i.e. in cryptocurrencies), such commitments may allow third parties to verify that the input and output amounts of a transaction are equivalent without revealing them. Other such cryptographic commitments are also being used - e.g. in the Ouroboros Crypsinous blockchain [92].

Last but not least, a blockchain mechanism called *Mimblewimble* (MW) should be also mentioned, which is a way of structuring and storing transactions - utilising a Pedersen commitment - so as to ensure that all transactions resemble random data to an outsider. The transaction data is only visible to their respective participants. Therefore, MW lies also in the concept of confidential transactions. Litecoin has announced the integration of MW, for enhancing privacy.

### 9.6. Homomorphic encryption

As stated in [26], this technique may be the vehicle to share and perform operations over data without revealing their private values, hence, implementing somehow the aforementioned idea of confidential transactions. Privacy mechanisms relying on homomorphic encryption are available in Ethereum and Quorum.

Due to their inherent properties, homomorphic ciphers allow arithmetic operations over the ciphertexts without having access to the decryption key, so as to ensure that the result coincides with the encrypted version of the value that we would get if the operations were initially performed over the original values. This is the case of the AZTEC protocol in the Ethereum (and also in some implementations of Quorum, as an enterprise-focused version of Ethereum), which utilises zero-knowledge proofs (being described next) as well as homomorphic encryption to encrypt the inputs and outputs of a transaction, so as to ensure that the blockchain can still test the logical correctness of these encrypted statements.

In general, there is a high potential for the integration of blockchain based-IoT with homomorphic encryption. The main task of this encryption in the IoT domain is securing IoT-Data with high privacy in decentralised mode. The task of this description is discussed in [170].

### 9.7. Zero-knowledge proofs

Towards implementing confidential transactions, the so-called zero-knowledge proofs (ZKP) can be also used. A ZKP is a cryptographic protocol with the following purpose: an entity (the prover) proves to another entity (the verifier) that she possesses/knows an information/statement, without exposing anything about the information/statement itself (e.g. it can be proved that she has at least 100 coins in her account, without revealing the actual amount). ZKPs are known in the cryptographic community for many decades [71]; they are being revisited though in the context of blockchains, being used by transacting parties to create confidential transactions, whose information can be verified without revealing the actual information. In this framework, the Zero-Knowledge Succinct Non-Interactive Argument of Knowledge (known as zk-SNARK) is being mainly used: *succinct* means that the zero-knowledge proofs can be verified within a few milliseconds, whilst *non-interactive* indicates that the proof consists of a single message sent from the prover (i.e. the claimant) to the verifier (and, thus, it is not an interactive protocol). In cases that zero knowledge proofs are implemented, the blockchain reveals only that a transaction has occurred, but not which public key transferred which amount to the recipient.

It has been pointed out that zero knowledge proofs and homomorphic encryption mentioned above have the potential to solve the conflict between data minimisation and the verifiability of data between many parties [149]. It is a highly evolving field, where several improvements of zk-SNARKs are being proposed - such as the less expensive implementation in [195].

Several blockchain platforms employ ZKPs, such as the Ethereum though the so-called AZTEC protocol and Quorum through the zero-knowledge security layer (ZSL) protocol. Ouroboros Crypsinous [92], which is built upon Ouroboros is also employing zk-SNARKs. The latest version Tron (v4.0) also introduces privacy techniques based on zk-SNARKs, by using relevant code from the well-known cryptocurrency Zcash. In Fabric, privacy-preserving authentication and transfer or certified attributes can be done using Identity Mixer (Idemix) which is a ZKP-based cryptographic protocol [36]. The developers of Stellar have introduced a zero-knowledge virtual machine being called ZkVM, allowing individuals and organisations performing safely their transactions directly on the shared ledger, instead of keeping them in siloed databases. Corda documents also refer to the intention on migrating to ZKP [154], whilst very recently the relevant idea seems to be implemented by utilising Corda in the bank sector [98]. Finally, according to the public information from the Tezos community [184], the Tezos developer community has been particularly interested in enabling private transactions by implementing zk-SNARKs, also based on the relevant implementation of the Zcash. It should be also stressed, as a last remark, that the Zerocoin project investigates how ZKPs can alleviate the privacy issues of Bitcoin.

### 9.8. Blockchain segregation

The notion of blockchain segregation can also be considered as an important privacy enhancing technology, which seems to be prominent for IoT applications; by these means, there is no shared ledger, with all the data (transactions) available to all participants; instead, each participant has access only to a subset of all transactions (let's say in a subset of the whole blockchain). Hence, unauthorised third parties cannot have any knowledge on the existence of transactions for which they should not be aware of.

Privacy mechanisms relying on blockchain segregation are available in several platforms, namely Corda, Hyperledger Fabric, Hyperledger Iroha and EOSIO. In Corda, only a set of predefined and identified participants can be part of a particular communication, while all the other participants in the network remain unaware of the transaction. However, a network service, being called notary, is being used which gets transaction information for several purposes (such as to ensure that double-spending does not occur in the case of cryptocurrency).

The Hyperledger Fabric supports two mechanisms for such a segregation. The first one rests with the so-called channels concept, which allows to separate the information between different channels [10]. The second one is the concept of the private data, which allows to isolate data between different organisations within the same channel. The channels are associated with the network level, whereas the private data are associated with the chaincode (application) level. Such privacy mechanisms are very important in cases of providing blockchain network and applications into a consortium environment (where consortium is a number of organisations with common business goals).

In the Hyperledger Iroha, access control rules along with some encryption maintain such a segregation. Access control rules can be defined at three levels: user-level, domain-level or system-level. At the user-level privacy rules for a specific individual are defined. If access rules are determined at domain or system level, they are affecting all users in the domain. In general, a role-based access control design has been implemented, where each role has specific permissions. In a somehow similar underlying logic (although fully different in the implementations), EOSIO also supports several permission schemes for the users; each permission is linked to an authority table which contains a threshold that must be reached in order to allow the action associated with the given permission to be executed.

### 9.9. Off-chain transactions/storage

By means of off-chain transactions, we refer to the case of using off-ledger channels to perform transactions, without broadcasting each transaction to the entire network. For example, off-ledger payment channel setups are implemented on public blockchains such as Bitcoin and Ethereum (Lightning and Raiden networks, respectively). This technique can possibly be used in all the blockchain platforms with proper implementation.

Generally we may decide, regardless the underlying blockchain platform, to not store all transactional data on the blockchain itself - and this is something that is expected to occur in IoT environments, due to the personal data protection issues that arise. Rather, such data could be stored in another, off-chain database and merely linked to the blockchain through a hash. Clearly, this process has a number of advantages from a data protection perspective.

For example, in the Hedera blockchain, based on the info in [40], the personal data is encrypted before being submitted to the Hedera Consensus Service. It will not be persisted on the main net nodes and will be persisted in that encrypted form on whichever mirror network nodes keep the messages. The mirror network nodes do not have the relevant decryption keys; these are being held only by the business application network. Also in the Ontology blockchain white paper, there is an explicit reference to the fact that the network structure supports both blockchain ledger storage and off-chain decentralised storage [144]. Exonum has been also used under this principle; for example, in [101] an Exonum-based health data ecosystem is described, divided into two segments — open and closed. The medical data is stored in the closed segment, whereas each patient unique identifier is stored in open segment.

### 9.10. Editable blockchains

The inherent immutability property of the blockchains results in a data protection concern related to the so-called right to be forgotten: data cannot be deleted. Therefore, since from a data protection perspective, there generally exists the right to data erasure which should be satisfied (e.g. if the user's data have been put to the ledger based on her explicit consent, then this consent should be able to be revoked if no other legal basis for maintaining them to the ledger exists), the fulfilment of such a right becomes an important challenge in blockchain platforms. Although there exist exemptions for this right, mainly depending on the legal basis of the processing or other specific circumstances (e.g. the GDPR provides such an exemption in cases that the data are necessary for the establishment, exercise or defence of legal claims), one should always consider how to handle the case that personal data are to be deleted, if this option may be in place.

Whenever such a data protection right is to be satisfied, a typical approach is to appropriately design the blockchain application so as to avoid storing plaintext data in the ledger but only to an off-line database (see subsection 9.9), whereas the blockchain stores an irreversible commitment (e.g. a hash value) of these data. Hence, deleting the data from the off-line database does not allow any subsequent restoration, since the commitment stored in the immutable ledger is irreversible. Therefore, the aforementioned usage of off-line database in subsection 9.9 also suffices to practically address this privacy concern (although, there still exist issues with respect to whether keeping only a commitment of the original data is always legally equivalent to deletion of data). However, the research community has started exploring the so-called editable blockchains (also known as redactable blockchains), using chameleon hash functions to edit, rewrite or delete data (see, e.g., [13, 74]). Chameleon (or trapdoor) hash functions allow an authorised party to compute a collision with a given hash value, whilst such collisions cannot be found by anyone else if this trapdoor is not known. It has, however, been stressed [149], that since a blockchain becomes editable, its initial advantages of using this solution compared to other forms of databases may be questionable. On the other side, such a solution may be preferable in cases that it is desirable, while preserving the tampering detection property, to explicitly allow someone with higher privileged access like an appropriate authority to be able to change the data or erase it - and this could possibly be the case of a private blockchain for IoT applications [74]. In any case, this is for the moment a new evolving area, not being implemented yet by the main stakeholders in the field.

### 9.11. Synopsis

Table 6, which summarises the main properties of the discussed blockchain platforms, contains also information on the main privacy mechanisms that are being employed in each of them; each mechanism is associated therein with its corresponding number of the subsection that is being described. Note that the pseudonymisation feature (subsection 9.1) is present to all platforms. Moreover, the off-chain storage (subsection 9.9) is also present to any platform supporting private permissioned solutions (since we assume that any such platform may easily implement this privacy feature, if necessary); however, it should pointed out that this feature could be possible implemented in other platforms too (see, e.g., the relevant discussion for Bitcoin and Ethereum in subsection 9.9). Finally, in cases that a blockchain platform provider has announced the adoption of a privacy mechanism in the near future, this is also included in Table 6.

We should stress that the number of adopted privacy mechanisms does not necessarily yields by itself a conclusion on whether privacy is present or not. As it is discussed earlier, depending on the context that a blockchain is being used, some privacy features may be unnecessary in some cases,

**Table 6**
Comparison of blockchain platforms in terms of security and privacy, PQC capabilities, as well as other characteristics

| Blockchain platforms | Type | Industry focus | Smart contracts language | Platform's security | PQC | Privacy mechanisms |
|---|---|---|---|---|---|---|
| Bitcoin | PL | Financial | Bitcoin script, Miniscript | Low (0.1) | ✓ | 9.1–9.3 |
| Bitcoin-NG | PL | Financial | – | Low (0.29) | – | 9.1–9.3 |
| Litecoin | PL | Financial | C++ | Low (0.29) | – | 9.1, 9.5 |
| Ethereum | PL | Cross-industry | Ethereum bytecode, LLL, Solidity | Low (0.27) | – | 9.1– 9.3, 9.5–9.7 |
| Elastico | PL | Financial | – | Med (0.36) | – | 9.1 |
| Parity | P | Cross-industry | Rust | Med (0.34) | – | 9.1, 9.5, 9.9 |
| Corda | P | Financial | Kotlin, Java | Med (0.49) | ✓ | 9.1, 9.7–9.9 |
| Stellar | PL/P | Financial | JavaScript, Golang, Java, Ruby, Python, C# | Med (0.42) | – | 9.1, 9.7, 9.9 |
| Fabric | P | Cross-industry | Golang, Java, Node.js | Med (0.45) | /[1] | 9.1, 9.7–9.9 |
| Symbiont | P | Financial | Symbiont Programming Language (SymPL) | Med (0.38) | – | 9.1, 9.5, 9.9 |
| Sawtooth | P | Cross-industry | C++, Go, Java, Solidity, Python, Rust, JavaScript | Low (0.28) | – | 9.1, 9.5, 9.9 |
| SMaRtChain | P | Cross-industry | – | Med (0.45) | – | 9.1, 9.9 |
| NEO | P | Smart economy | C#, VB.Net, Kotlin, F#, Python, JavaScript | Med (0.43) | / | 9.1, 9.9 |
| EOSIO | P | Cross-industry | C, C++, Wasm | Med (0.31) | – | 9.1, 9.8, 9.9 |
| Bitshares | PL | Financial | C++ | Med (0.36) | – | 9.1, 9.5 |
| Tron | PL | Financial | Solidity | Med (0.42) | – | 9.1, 9.7 |
| Lisk | PL | Financial | JavaScript, TypeScript | Med (0.32) | – | 9.1 |
| Tezos | PL | Cross-industry | Michelson | Med (0.54) | – | 9.1, 9.7 |
| Exonum | P | Cross-industry | Java, Rust | Med (0.64) | – | 9.1, 9.9 |
| Quorum | P | Cross-industry | Solidity, Vyper | Med (0.49) | – | 9.1, 9.5–9.7, 9.9 |
| Ripple | P | Financial | Any language | Med (0.48) | – | 9.1, 9.2, 9.9 |
| Ethermint | PL/P | Financial | Ethereum smart contracts | Med (0.4) | – | 9.1, 9.9 |
| Burrow | P | Cross-industry | Ethereum smart contracts Web Assembly (WASM) | Med (0.45) | – | 9.1, 9.9 |
| Ontology | P | Cross-industry | C, Python | Med (0.42) | – | 9.1, 9.9 |
| Iroha | P | Cross-industry | Native | Med (0.39) | – | 9.1, 9.8, 9.9 |
| IOTA | PL | IoT | Qupla, Abra VM | Med (0.54) | ✓ | 9.1, 9.3, 9.9 |
| Hedera | P | Cross-industry | Solidity, Vyper | High (0.74) | – | 9.1, 9.9 |
| Ouroboros | PL | Financial | – | Med (0.4) | – | 9.1, 9.5, 9.7 |
| Snow-White | P | Financial | – | Med (0.39) | – | 9.1, 9.9 |
| Algorand | PL | Financial | TEAL, Clarity | Med (0.39) | / | 9.1, 9.9 |
| Monero | PL | Financial | –[2] | Med (0.39) | – | 9.1, 9.2, 9.4 |
| Besu | P | Cross-industry | Solidity, Vyper | Med (0.42) | – | 9.1, 9.5, 9.9 |

✓ denotes that techniques against cryptanalitic attacks have being implemented;
/ denotes that the platform is transitioning to a quantum-resistant infrastructure;
− denotes that the factor has not yet being implemented;
[1] PQ-Fabric [82] is not yet peer-reviewed and therefore not considered for our evaluation.
[2] Tari [181], written in Rust, can be used as a secondary layer to provide the smart contracts' functionality in Monero.

as well as, on the contrary, insufficient in some other cases. Therefore, Table 6 simply surveys the adopted privacy mechanisms per blockchain platform and should not be used to evaluate these platforms in terms of privacy.

Concluding, it should be pointed out that, in general, future developments of blockchains may in turn pose new privacy threats, thus rendering their overall evaluation a challenging and evolving task. For example, according to a statement on the Hedera's website (which currently employs permissioned nodes), the plan is to include permissionless nodes in the future, towards growing the network. Such an extension clearly affects the privacy requirements. Therefore, in such changes of a platform, its applicability in a SOHO environment should be considered on a new basis. In other words, towards examining the blockchain platforms, a long-term approach needs to be followed - i.e. it is essential to also take into account the planned changes of the platform.

# 10. Blockchains' comparative evaluation

The consensus protocols discussed in section 6 have been implemented into various blockchain platforms, with different benefits, shortcomings and security levels. The adversarial tolerance of each protocol, along with its performance; i.e. scalability, transactions' efficiency and block confirmation time are considered to be the most critical factors in an IoT implementation. Therefore, it is compulsory each transaction that contains IoT data to be stored to a block in a matter of seconds or even less [131]. As already mentioned, throughput, latency, scalability are mapped to a three-valued scale: *low*, *medium*, and *high*. As to the terms of performance, even with the same scalable and secure consensus protocol, slightly changing some parameters of its design, can result into different performance metrics. Therefore, in this section, we compare the aforementioned solutions for the needs of a blockchain-enabled smart home ecosystem.

Despite PoW's security advantages, several defects and restraints of PoW-based consensus algorithms exist and are becoming more and more evident. Such protocols have many restraints that IoT devices cannot solve. The mining process requires high computational power and high bandwidth to establish consensus, while most of the IoT devices posses low processing capabilities and elementary hardware specifications with very low storage capabilities. Thus, the resource demanding mining tasks cannot be executed on low power home devices and the requirements that are needed to store an entire blockchain mechanism cannot be satisfied.

PoS and its variants can significantly reduce the energy consumption that PoW-based consensus protocols introduce and allow IoT nodes to participate in the consensus process. Despite of this, new security issues might arise that cannot be easily solved in a blockchain-based smart home ecosystem. The nodes with the highest amounts of stakes, have higher probability to create the next block; an issue that does not offer much in terms of decentralisation and makes the protocol somehow centralised with only a few number nodes to be able to create a new block. Although Ouroboros [94] and Snow White [25] are both PoS consensus protocols, they present interesting differences between them in the leader election process. In Ouroboros, a subset of stakeholders executes a multi-party coin tossing protocol and agrees upon a random seed that is given into a pseudo-random function that outputs a leader according to its accumulated stake. This seed is also used to define the next subset of stakeholders in the following round and the reward for the creation of the block to be distributed between all the subset's members. On the other hand, Snow White is based on a weakly synchronised clock and employs a private leader election process. Nevertheless, PoS protocols cannot provide the necessary performance requirements to be deployed in an smart home ecosystem, as shown in Table 7 and they are prone to various attacks. In a "nothing-at-stake-attack", an adversary has nothing to loose, if it starts to behave maliciously and create alternative chains. This attack could not be easily addressed in the referenced IoT ecosystem.

In the case of hybrid consensus protocols (PoW and BFT or in PoS and BA), the voting weight of each node is increased according to its block proposing capabilities. This augmentation is established either by linking the computational capabilities of each node to its opportunity of participating in the BFT consensus process (e.g. Elastico) or by weighting each nodes's vote to its stake. Algorand guarantees that the probability of forks is negligible, which is a significant advantage over the classical PoS protocols, since various attacks can be mitigated, such as the aforementioned nothing-at-stake attack. The protocol's design is such that it allows to scale to millions of users and maintain a high transaction rate, which is a significant appealing feature for any IoT implementation.

Other hybrid consensus protocols, like PoA and PoET-SGX, utilise the computational power for identity management and therefore the fault tolerance threshold is based on the percentage of the malicious identities. DAG-based, such as Tangle achieve probabilistic finality, high scalability and performance, as well as, 50% adversarial tolerance of computing power. IOTA is designed to meet the necessary requirements for the integration of the blockchain technology to the IoT.

PBFT-based consensus protocols cannot be directly implemented to public and permissionless settings. However, higher throughput and lower latency can be achieved, in contrast to protocols implemented to permissionless settings, such as PoW and PoS. The fixed-size group of VRs/delegates in PBFT and DPoS, in which less $1/3$ can be malicious to establish agreement for the translations' total order, is a factor that significantly impacts their nomination for a smart home implementation, despite their high performance metrics.

IBFT provides immediate block finality, as all the deterministic protocols do, high data integrity, byzantine fault tolerance and high level of governance. Unfortunately, the protocol lacks in terms of scalability, since it does not scale well under a totally decentralised network, with the IoT devices that participate in the consensus process to be fixed and difficult to be modified.

RPCA is designed for financial purposes and requires less than $1/5$ of the nodes to be faulty in the UNL to ensure network consensus. Therefore, when a specific level of trust is established, the protocol seems to trade adversarial tolerance with better performance metrics. On the down side, the multi-round dissemination scheme between each UNL, as well as, the requirements for quick convergence of votes, demand from to protocol to be highly synchronous; a requirement does not provide much to RPCA's decentralisation.

Kafka and Raft are CFT consensus protocols, which are mostly known for their high performance metrics in Fabric. These protocols cannot withstand malicious behaviour. For this reason, Kafka and Raft are implemented in permissioned blockchain platforms (Corda and Fabric) that do not follow the order-execute design, in which the transactions are ordered first and then executed in the same order on all the nodes sequentially. On the contrary, these two platforms introduce the execute-order-validate architecture, in which

**Table 7**
Comparison of blockchain consensus protocols, in terms of fault tolerance and performance requirements, also incorporating the security and privacy requirements illustrated in Table 6, for the needs of the IoT applications under consideration.

| Consensus protocol | Blockchain platforms | Fault tolerance | Scalability | TPS | BCT | Overall score | SOHO suitability |
|---|---|---|---|---|---|---|---|
| PoW | Bitcoin | < 50% comput. power | Low | 7 | 10min | 37% | ◐ |
| | Litecoin | | | 56 | 2.5min | 34% | ◐ |
| | Monero | | | 1.7K | 2min | 49% | ◐ |
| | Ethereum | | | 25 | 10−15sec | 34% | ◐ |
| PoW & BFT | Bitcoin-NG | < 50% comput. power | Low | 200 | 10min | 36% | ◐ |
| | Elastico | BFT: $4f+1$ | | 40 | 110sec | 35% | ◐ |
| PoS | Ouroboros | < 50% stake | Low | 257 | 2min | 38% | ◐ |
| | Snow-White | | | 100−150 | 10min | 42% | ◐ |
| PoS & BA | Algorand | BFT: $3f+1$ [1] | High | 1K | 5sec | 68% | ◐ |
| DpoS | EOSIO | | | 1K−6K | < 1sec | 65% | ◐ |
| | Bitshares | | | 100K | 1sec | 60% | ◐ |
| | Lisk | BFT: $3f+1$ [2] | Low | 2.5 | 6min | 35% | ◐ |
| | Tron | | | > 2K | 3sec | 58% | ◐ |
| | Tezos | | | ≈ 40 | ≈ 30min | 38% | ◐ |
| Kafka | Fabric | CFT: $2t+1$ | Medium | 3.5K/20K | < 1sec | 67% | ◐ |
| Raft | Corda | | | 170 | 1sec | 57% | ◐ |
| | Quorum | CFT: $2t+1$ | Medium | ≈ 650 | 4.5sec | 53% | ◐ |
| | Fabric | | | ≈ 7K | < 1sec | 68% | ◐ |
| IBFT | Quorum | BFT: $3f+1$ | Low | ≈ 600 | 4.5sec | 53% | ◐ |
| | Besu | | | 350−400 | 2sec | 52% | ◐ |
| Aura | Parity | BFT: $2f+1$ | High | 35−45 | 5sec | 65% | ◐ |
| BFT-SMaRt | Fabric | BFT: $3f+1$ | High | > 10K | 0.5sec | 87% | ● |
| | Symbiont | | | 80K | < 1sec | 86% | ● |
| | SMaRtChain | | | 12K−13K | < 0.25sec | 87% | ● |
| VBFT | Ontology | BFT: $3f+1$ | High | 3K−4K | 5sec | 81% | ● |
| DBFT | NEO | BFT: $3f+1$ | High | < 1K | 15−20sec | 69% | ◐ |
| Tendermint | Ethermint | BFT: $3f+1$ | Medium | 200−800 | < 1sec | 67% | ◐ |
| | Burrow | | | > 200 | 2sec | 61% | ◐ |
| Exonum | Exonum | BFT: $3f+1$ | Medium | ≈ 5K | 0.5sec | 80% | ● |
| PoET-SGX | Sawtooth | BFT: $\Theta(\log \log n / \log n)$ | High | 1K−2.3K | < 1sec | 84% | ● |
| Ripple | Ripple | BFT: $5f+1$ [3] | High | 1.5K | 4sec | 79% | ● |
| YAC | Iroha | BFT: $3f+1$ | Medium | × 1K | < 3sec | 74% | ● |
| SCP | Stellar | BFT: $3f+1$ | High | 1K | 3sec | 76% | ● |
| Tangle | IOTA | < 50% comput. power | High | 1.5K | 10ms | 84% | ● |
| Hashgraph | Hedera | BFT: $3f+1$ | High | 10K | 5sec | 86% | ● |

[1] of maliciously-possessed stakes
[2] of delegates;
[3] of nodes in each UNL

the transactions are executed first; for their validity to be checked (stateless validation), ordered via the consensus mechanism and then validated again (stateful validation) with all the transactions to be included to the ledger (either valid or invalid). Such a modification to the ordinary transaction flow, might seems to "give room" to potential adversaries to take advantage of the CFT consensus process and insert malicious transactions on the ledger. On the contrary; the validation of the transactions is occurring only in the execution and validation phase [9], while in the consensus process the nodes only establish their total order without any validation checks. In contrast to BFT protocols, in which the validation replicas already vote to elect the leader, a CFT implementation in Fabric can not only accommodate flexible trust and fault assumptions, but also to separate the trust scheme for applications from the trust scheme of the consensus process; and therefore, it can allow to store trustworthy IoT data to the distributed ledger [2]. Various other cases also exist that have implemented Kafka or Raft in Fabric to address the needs of the IoT [38, 96].

PoW, PoS and their variations, as well as, DPoS, Aura, PoET, Tangle and Hashgraph are probabilistic protocols and therefore, in these protocols forks can easily arise; in contrast to the rest that are deterministic. For example, IOTA and Hedera are both currently projects using consensus protocols based on *directed acyclic graphs* (DAGs) to provide high transactions' throughput and low BCT. The transaction's finality in IOTA is probabilistic and it is based on a *Markov Chain Monte Carlo* (MCMC) method, while the finality in Hedera is deterministic, meaning that if forks occur (in probabilistic protocols), they cannot be easily addressed in a network with resource constrained IoT nodes. On the other hand, DAGs become more and more wide, with high performance metrics and they are not limited to linear processing, as observed in the most blockchain platforms. Therefore, DAGs can be considered to be the solution for the scalability issue in any IoT ecosystem.

Permissionless blockchain platforms are affected by the propagation speed in the network and hence provide lower performance metrics and scalability. On the other hand, permissioned blockchain platforms have much better overall performance, but implement voting mechanisms and introduce network overhead limiting the system's scalability, with the worst case complexity to be $\mathcal{O}(N^2)$ compared to that of permissionless, which is $\mathcal{O}(N)$. Therefore, the suitability of permissioned blockchain implementations to the IoT communication is limited by each protocol's communication complexity. Via the virtues of public, anonymised access and decentralisation, permissionless blockchain platforms are more applicable to industry-wide IoT applications, while the permissioned are more applicable to enterprise-based solutions, due to their significantly higher degree of access control and permission allocating capabilities.

In contrast to other works, where the corresponding suitability of each consensus protocol seems to be vague, we define the suitability of the consensus protocol $i$ and the blockchain platform $j$ to the referenced IoT ecosystem (smart home) as follows

$$\text{soho}(i,j) = w_1 \, \text{pft}_i + w_2 \, \text{scl}_i + w_3 \, \text{sec}_j + w_4 \, \text{prv}_{i,j} \\ + w_5 \, \text{pqc}_j + w_6 \, \text{thr}_{i,j} + w_7 \, \text{bct}_{i,j} \quad (5)$$

where, each one of the corresponding parameters are defined subsequently. The fault tolerance $\text{pft}_i$ of the consensus protocol $i$ is defined as

$$\text{pft}_i = \begin{cases} 1, & \text{if } i \text{ tolerates adversarial behavior,} \\ 0.5, & \text{if } i \text{ tolerates crashes only,} \\ 0, & \text{otherwise (i.e. if } i \text{ is centralised).} \end{cases} \quad (6)$$

Clearly, centralised consensus protocols, such as the Solo protocol in Fabric, the single node protocol in Corda, etc., are not suggested for any IoT application. Each of these protocols, creates a single point of failure and therefore, such protocols are not included in our work. The scalability $\text{scl}_i$ of consensus protocol $i$ equals

$$\text{scl}_i = \begin{cases} 1, & \text{if } i\text{'s scalability is High,} \\ 0.5, & \text{if } i\text{'s scalability is Medium,} \\ 0, & \text{if } i\text{'s scalability is Low} \end{cases} \quad (7)$$

whereas the security $\text{sec}_j$ of blockchain platform $j$ is given by (1). The privacy that the combination of the consensus protocol $i$ and blockchain platform $j$ provides, is denoted by $\text{prv}_{i,j}$. As shown in section 9 and Table 6, all platforms and their consensus protocols considered in this work support a sufficient number of strong privacy-preserving mechanisms. Hence, it can be seen that the type of the blockchain platform (i.e. whether $j$ is permissioned or permissionless) should be the dominant factor for evaluating privacy; this is reflected to the following expression

$$\text{prv}_{i,j} = \begin{cases} 1, & \text{if } j \text{ is permissioned,} \\ 0.5, & \text{otherwise} \end{cases} \quad (8)$$

due to the fact that permissionless blockchain platforms have in principle higher privacy risks compared to permissioned, which are more capable to satisfy the necessary flavors of privacy in an smart home ecosystem. The capability $\text{pqc}_j$ of the blockchain platform $j$ to implement post quantum signatures and post-quantum cryptography (PQC) in general, leads to

$$\text{pqc}_j = \begin{cases} 1, & \text{if } j \text{ implements PQC,} \\ 0.5, & \text{otherwise} \end{cases} \quad (9)$$

since apart from the implementation of PQC schemes, other factors, like cryptographic keys' sizes, also play an important role towards withstanding various cryptographic attacks, e.g. brute force attacks in Table 2. Based on the analysis presented in section 4.1, the transactions' throughput $\text{thr}_{i,j}$ that consensus protocol $i$ can achieve when implemented in the blockchain platform $j$ is defined as

$$\text{thr}_{i,j} = \min\left(0.05 \cdot \lfloor \text{TPS}_{i,j}/100 \rfloor, 1\right). \quad (10)$$

The block confirmation time $\text{bct}_{i,j}$ that the consensus protocol $i$ can achieve when implemented in the blockchain platform $j$ is likewise given by the expression

$$\text{bct}_{i,j} = \max\left(1 - 0.1 \cdot \lfloor \text{BCT}_{i,j} \rfloor, 0\right). \quad (11)$$

If the $\text{BCT}_{i,j}$ is greater than 10sec, then it is considered as unsuitable for a sustainable smart home blockchain ecosystem. The weights given to (5) are equal to $w_1 = w_2 = 0.2$, $w_3 = w_6 = 0.15$, and $w_4 = w_5 = w_7 = 0.1$, with

$$w_1 + w_2 + \cdots + w_7 = 1. \quad (12)$$

The assigned weights are categorised and separated into security (55%) and performance (45%); and they are selected in a way so as to give more weight to security since any IoT ecosystem is clearly an adversarial environment. Nevertheless, for the aforementioned consensus protocols and blockchain platforms, we have provided additional information, such as, consensus finality, industry focus – monetary concepts, smart contracts' implementations and privacy preservation mechanisms. In our work, a blockchain-enabled smart home implementation is mostly focused only on the critical features of security, privacy and performance that the blockchain technology should provide to any IoT application, if used. Therefore, the assigned weights are mostly given to the consensus protocol's security, scalability and performance (a total weight of 65%), since, this is where the IoT nodes

Blockchains for the IoT: architectures, security, privacy, and performanceface the most significant challenges. The rest (35%) is given to the platform's security and privacy.

As illustrated in Table 7, by taking into consideration all the established evaluation criteria (related to the platforms' security and privacy, but also to the underlying protocols' performance, fault tolerance and scalability) only six consensus protocols out of sixteen can achieve high suitability score so as to be implemented to the smart home ecosystem; these are BFT-SMaRt, PoET, VBFT, Exonum, Tangle, and Hashgraph.

**Example 3.** When BFT-SMaRt is implemented in Fabric, Symbiont or SMaRtChain, it can achieve high suitability scores of 87%, 86% and 87%, respectively. The consensus protocol is secure against malicious behaviour, scalable and when it is implemented in Fabric, high throughput and low block confirmation time can be achieved. Fabric is a permissioned ledger and achieves a medium level of security, while techniques against cryptanalytic attacks have not yet being adopted. Therefore, for $i = 1$, from the equation 5, we have

$$\begin{aligned}\mathsf{soho}(1,2) &= w_1\,\mathsf{pft}_1 + w_2\,\mathsf{scl}_1 + w_3\,\mathsf{sec}_2 + w_4\,\mathsf{prv}_{1,2} \\ &\quad + w_5\,\mathsf{pqc}_2 + w_6\,\mathsf{thr}_{1,2} + w_7\,\mathsf{bct}_{1,2} \\ &= 0.2 \cdot 1 + 0.2 \cdot 1 + 0.15 \cdot 0.45 + 0.1 \cdot 1 \\ &\quad + 0.1 \cdot 0.5 + 0.15 \cdot 1 + 0.1 \cdot 1 \approx 0.87\end{aligned}$$

the overall SOHO suitability score of BFT-SMaRt, when it is implemented in Fabric, to be $\approx 87\%$. ∎

On the other hand, as it was anticipated, the cryptocurrency based ledgers and the protocols that rely on PoW or PoS to establish consensus, receive low suitability scores. Without taking into consideration the computational power required to establish consensus, which is an important factor for any blockchain-based IoT implementation, these protocols receive scores bellow 50%. Therefore, the computational overhead is not the only factor that makes the traditional blockchain solutions unsuitable for the needs of the IoT. The only exception from these platforms is Algorand, which receives the overall suitability score of 68%.

**Example 4.** To be more detailed, Algorand relies on PoS integrated in a BA layout to achieve optimal resilience against faulty nodes and high scalability. Algorand is a permissionless platform and achieves a medium level of security, while techniques against cryptanalytic attacks have not yet being adopted. Algorand can also achieve throughput of 1000 TPS with each transaction to be confirmed in 5sec. Therefore, for $i = 2$, Algorand's suitability to a SOHO ecosystem is defined from equation 5, and it is

$$\begin{aligned}\mathsf{soho}(2,1) &= w_1\,\mathsf{pft}_2 + w_2\,\mathsf{scl}_2 + w_3\,\mathsf{sec}_1 + w_4\,\mathsf{prv}_{2,1} \\ &\quad + w_5\,\mathsf{pqc}_1 + w_6\,\mathsf{thr}_{2,1} + w_7\,\mathsf{bct}_{2,1} \\ &= 0.2 \cdot 1 + 0.2 \cdot 1 + 0.15 \cdot 0.39 + 0.1 \cdot 0.5 \\ &\quad + 0.1 \cdot 0.5 + 0.15 \cdot 0.5 + 0.1 \cdot 0.5 \approx 0.68\end{aligned}$$

and thus the overall SOHO suitability is 0.68. ∎

Apart from DPoS implementation in Lisk, the rest of the consensus protocols, in any possible implementation, achieve medium suitability scores between 50% and 80%, with Ripple to achieve the highest score among them (79%). However, our evaluation framework can be used with different weights and impose other quantitative criteria to select the best solution for different IoT applications, according to each specific use case.

## 11. Conclusions

This paper conducted a comprehensive and coherent review of the available blockchain solutions to determine their ability to meet the requirements and address the challenges of the IoT ecosystem. Having the smart home ecosystem as our reference IoT domain, a number of requirements were extracted so as to assess the strengths and shortcomings of the investigated blockchains when utilised as a building block to deliver relevant applications and services to the end users. A large number of characteristics were considered during our study, covering various architectural components of blockchain solutions, that are classified into the broad areas of

- *Security*: protocols' fault tolerance, network security, overall platforms' security including smart contracts, and cryptographic strength;
- *Privacy*: both well-known but also emerging privacy-preserving mechanisms were investigated; as well as
- *Performance*: throughput, block confirmation time / latency, and scalability.

Amongst the key findings of our work, was that the defences currently provided by blockchain platforms are not sufficient to thwart all the prominent attacks against blockchains, with blockchain 1.0 and 2.0 platforms being susceptible to the majority of them. This seems to be in contrast with privacy preservation mechanisms, which are embraced, to varying degrees, by all the platforms investigated (at a minimum supporting the features of pseudonymisation and off-chain storage). If the underlying consensus protocols' performance and fault tolerance is also considered, then only a small number of platforms meet the requirements of our reference IoT ecosystem (smart home).

This work provides a holistic approach that could however be further explored in different ways so as to yield valuable research or practical results. Such explorations could consider the accurate modelling of the blockchain platforms' consensus protocols, and other operations that play a key role into determining performance or storage criteria, so as to derive measurements based on the same underlying assumptions; this constitutes part of our ongoing research. At the practical side, this goal could also be achieved by performing in-lab measurements (for the platforms allowing to do so) using a given use case scenario, thus allowing to get results under the same operational conditions.

Brotsis, et al.: *Preprint submitted to Elsevier* Page 31 of 37

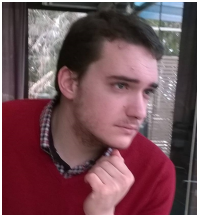
Sotirios Brotsis received the B.S. degree from the Department of Mathematics, Greece, University of Patras and the M.S. degree from the Department of Informatics and Telecommunications, Greece, University of Peloponnese. He is currently pursuing the Ph.D. degree with the same department, supervised by the Associate Professor N. Kolokotronis. His research interests lie in Blockchain technology, cryptography and IoT security. He has published his research in top venues, and he has been a TPC member and a regular reviewer for international journals and conferences.

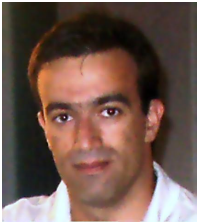
Konstantinos Limniotis received the Ph.D. in cryptography from the National and Kapodistrian University of Athens in 2007. He is currently an ICT Specialist at the Hellenic Data Protection Authority. He also holds visiting positions, as an Adjunct Faculty Member, at the Open University of Cyprus and the National and Kapodistrian University of Athens, Greece. In addition, he is a research associate with the University of Peloponnese, Greece, since 2017. His main research interests include cryptography, personal data protection and information security. He has over 40 publications in these areas, whilst he is regular reviewer in international journals and conferences.

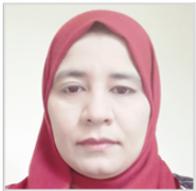
Guletoum Bendiab has a PhD degree in Computer Science. Her research interests are in the areas of trust and identity management, Blockchain, machine learning with applications in cyber security and most specifically in digital forensics, Intrusion Prevention Detection and Response, Social Engineering, OSINT, Insider Threats, Malware and Distributed Denial of Service Attacks.

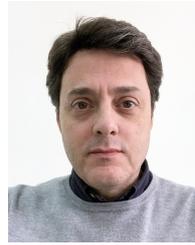
Nicholas Kolokotronis is an Associate Professor and head of the Cryptography and Security Group at the Department of Informatics and Telecommunications, University of the Peloponnese. He received his B.Sc. in mathematics from the Aristotle University of Thessaloniki, Greece (1995), M.Sc. in highly efficient algorithms (1998, highest honors) and Ph.D. in cryptography (2003) from the University of Athens. He has been a cochair and member of the organising committee in many international conferences. His research interests span cryptography, security, and coding theory, areas in which he has published more than 85 papers in international scientific journals, conferences, and books.

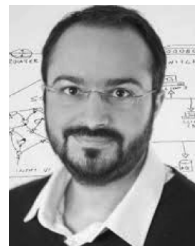
Stavros Shiaeles is currently Senior Lecturer at University of Portsmouth, UK. He holds various professional certifications in Cyber Security and he is an EC-Council Accredited Instructor, delivering CEH to professionals. He has published more than 70 articles in international scientific journals, conferences and books, he has been a TPC member and a regular reviewer for a number of international journals and conferences, and he is actively involved with EU-funded and national research and development projects. His research interests span the broad areas of cybersecurity, open-source intelligence, trust, blockchain, digital forensics, and machine learning applications in cybersecurity.